\documentclass{article}
\usepackage{graphicx}
\usepackage{amsmath}
\usepackage{amsbsy}
\usepackage{epstopdf, epsfig}
\usepackage[a4paper, total={6in, 8in}]{geometry}
\usepackage{longtable}
\usepackage{newtxtext,newtxmath}

\title{An upscaled model for permeable biofilm in a thin channel and tube}
\author{D. Landa-Marb\'an$^{1}$ \and
  G. B\o dtker$^{1}$ \and
  K. Kumar$^{2}$ \and
  I. S. Pop$^{3,4}$ \and
 \and F. A. Radu$^{4}$}
\date{}
\begin{document}
\maketitle
\noindent ${}^1$ NORCE, Nyg{\aa}rdsgaten 112, 5008 Bergen, Norway.\\[5pt]
${}^2$ Department of Mathematics and Computer Science, Karlstad University, Universitetsgatan 2, 651 88 Karlstad, Sweden.\\[5pt]
${}^3$ Faculty of Sciences, Hasselt University, Campus Diepenbeek, Agoralaan building D, BE3590 Diepenbeek, Belgium.\\[5pt]
${}^4$ Department of Mathematics, Faculty of Mathematics and Natural Sciences, University of Bergen, All\'egaten 41, P.O. Box 7803, 5020 Bergen, Norway.\\[5pt]
Corresponding author: David Landa-Marb\'an (E-mail: dmar@norceresearch.no)
\begin{abstract}
\noindent In this paper, we derive upscaled equations for modelling biofilm growth in porous media. The resulting macro-scale mathematical models consider permeable multi-species biofilm including water flow, transport, detachment and reactions. The biofilm is composed of extracellular polymeric substances (EPS), water, active bacteria and dead bacteria. The free flow is described by the Stokes and continuity equations and the water flux inside the biofilm by the Brinkman and continuity equations. The nutrients are transported in the water phase by convection and diffusion. This pore-scale model includes variations of the biofilm composition and size due to reproduction of bacteria, production of EPS, death of bacteria and shear forces. The model includes a water-biofilm interface between the free flow and the biofilm. Homogenization techniques are applied to obtain upscaled models in a thin channel and a tube, by investigating the limit as the ratio of the aperture to the length $\varepsilon$ of both geometries approaches to zero. As $\varepsilon$ gets smaller, we obtain that the percentage of biofilm coverage area over time predicted by the pore-scale model approaches the one obtained
using the effective equations, which shows a correspondence between both models. The two derived  porosity-permeability relations are compared to two empirical relations from the literature. The resulting numerical computations are presented to compare the outcome of the effective (upscaled) models for the two mentioned geometries. 
\end{abstract}

\newpage
\begin{longtable}{l l}
\textbf{List of Symbols}\\
$\mathbb{I}$&Identity matrix\\
$a$&Coverage area\\
$c$&Nutrient concentration\\
$B$&Relative porosity\\
$B_c$&Critical point\\
$d$&Biofilm height\\
$D$&Nutrient diffusion coefficient\\
$E$&Integration coefficient\\
$f^+$&Positive cut\\
$f^-$&Negative cut\\
$F$&Integration coefficient\\
$G$&Integration coefficient\\
$h$&Variable dependent on the biofilm height (channel)\\
$H$&Set-valued Heaviside graph\\
$H_\delta$&Regularized set-valued Heaviside graph\\
$i$&Imaginary number\\
$\pmb{J}$&Nutrient flux\\
$J_{\nu}$&Bessel function of order $\nu$ of first kind\\
$k$&Biofilm permeability\\
$K$&Permeability\\
$k_{res}$&Bacterial decay rate coefficient\\
$k_{str}$&Stress coefficient\\
$k_n$&Monod-half nutrient velocity coefficient\\
$l$&Half height of the channel\\
$L$&Channel/Tube length\\
$\mathbb{M}$&Matrix with flux water derivatives\\
$p$&Pressure\\
Pe&P\'eclet number\\
$\pmb{q}$&Water velocity\\
$\pmb{r}$& Vector (cylindrical coordinates)\\
$r$&Radial coordinate\\
$R$&Reaction term\\
$S$&Tangential shear stress\\
$t$&Time\\
$T$&Final time\\
$\pmb{u}$&Velocity of the biomass\\
$v$&Darcy velocity\\
$V$&Integration coefficient\\
$w$&Variable dependent on the biofilm height (tube)\\
$W$&Integration coefficient\\
$\pmb{x}$& Vector (Cartesian coordinates)\\
$x$&Cartesian coordinate\\
$X$&Integration coefficient\\
$y$&Cartesian coordinate\\
$Y$&Yield coefficient\\
$Y_{\nu}$&Bessel function of order $\nu$ of second kind\\
$z$&Cartesian/Cylindrical coordinate\\
\phantom{}\\
\textbf{Greek symbols}\\
$\chi$&General variable\\
$\delta$&Small regularization parameter\\
$\epsilon$&Dimensionless aspect ratio (channel)\\
$\varepsilon$&Dimensionless aspect ratio (tube)\\
$\eta$&Experimentally determined parameter\\
$\Gamma$&Space boundary\\
$\kappa$&Effective permeability\\
$\mu$&Water viscosity\\
$\mu_n$&Maximum rate of nutrient utilization\\
$\eta$&Experimental determined parameter\\
$\pmb{\nu}$&Unitary normal vector (interface)\\
$\nu_n$&Interface velocity\\
$\Omega$&Space domain\\
$\phi$&Porosity of porous media\\
$\varphi$&Angular coordinate\\
$\Phi$&Growth velocity potential\\
$\rho$&Density\\
$\varrho$&Tube radius\\
$\Sigma$&Sum of reaction terms\\
$\pmb{\tau}$&Unitary tangential vector\\
$\theta$&Volume fraction\\
$\pmb{\upsilon}$&Unitary normal vector (wall)\\
$\xi$&Variable dependent on permeability and water content (tube)\\
$\Xi$&Space region\\
$\zeta$&Tolerance\\
\textbf{Subscripts/superscripts}\\
$a$&Active bacteria \\
$B$&Biodegradation microbe\\
$b$&Biofilm\\
$C$&Channel\\
$c$&Critical\\
$d$&Dead bacteria\\
$i$&Input\\
$ib$&Input biofilm domain\\
$iw$&Input water domain\\
$K$&Biobarrier-forming microbe\\
$l$&Lower\\
$m$&Middle\\
$O$&Initial\\
$o$&Output\\
$ob$&Output biofilm domain\\
$ow$&Output water domain\\
$e$&EPS\\
$ref$&Reference value\\
$r$&r-component\\
$s$&Wall\\
$T$&Tube\\
$u$&Up\\
$w$&Water\\
$wb$&Water-biofilm (interface)\\
$y$&y-component\\
$z$&z-component\\
$0$&Lowest order term (asymptotic expansion)\\
$\tilde{}$&Dimensionless parameter/variable (channel)\\
$\bar{}$&Dimensionless parameter/variable (tube)\\
\textbf{Abbreviations}\\
EPS&Extracellular polymeric substance\\
MEOR&Microbial enhanced oil recovery\\
\end{longtable}

\section{Introduction}\label{intro}
Biofilms are sessile communities of bacteria housed in a self-produced adhesive matrix consisting of extracellular polymeric substances (EPS), including polysaccharides, proteins, lipids and DNA (\cite{Aggarwal:Article:2015}). The proportion of EPS in biofilms is 50$\%$ to 90$\%$ of the total organic matter (\cite{Donlan:Article:2002,Vu:Article:2009}). Water is by far the largest component of the matrix, giving biofilms the nickname `stiff water' (\cite{Flemming:Article:2010}). Biofilms provoke chronic bacterial infection, infection on medical devices, deterioration of water quality and the contamination of food (\cite{Kokare:Article:2009}). On the other hand, biofilms can be used for wastewater treatment and bioenergy production (\cite{Miranda:Article:2017}). In microbial enhanced oil recovery (MEOR), one of the strategies is selective plugging, where bacteria are used to form biofilm in highly permeable zones to diverge the water flow and extract the oil located in low-permeability zones (\cite{Raiders:Article:1989}). In wastewater treatment, one of the strategies consists of using biofilms to break down compounds which are not desirable to discharge into the natural environment (\cite{Capdeville:Article:1992}). 

Two of the motivations to derive upscaled models are to accurately describe the average behaviour of the system with relatively low computational effort compared to fully detailed calculations starting at the microscale (\cite{Noorden:Article:2010}) and to determine effective parameters (\cite{Helmig:Article:2002}). The values of these effective parameters can be determined using known values from pore-scale experiments. Recent works have been carried out to derive upscaled models, e.g., \cite{Collis:Article:2017} obtained a mathematical model describing macroscopic tumour growth, transport of drug and nutrient through homogenization and \cite{Jin:Article:2019} upscaled a pore-scale model for primary fluid recovery and showed that the macroscopic equation for the water flux is fundamentally different from Darcys' law. We also refer to \cite{Peszynska:Article:2016} and \cite{Schulz:Article:2017}, where the authors upscaled various pore-scale models for biofilm formation in perforated and strip geometries.

The present work builds on \cite{Landa:Article:2019}, where a pore-scale model is discussed. The model includes permeable biofilm and evolution of different biofilm components: active bacteria, dead bacteria and EPS. The importance of including biofilm
permeability is underlined by the fact that the dominated mechanism of nutrient transport within some biofilms is convection (\cite{Lewandowski:Article:2003}). This mathematical model is based on laboratory experiments performed by \cite{Liu:Article:2019}, where the biofilm was grown in micro-channels. Here we upscale this pore-scale model to derive effective equations, by investigating the limit as the ratio of the height to the length of the micro-channel approaches to zero.  

In this general context, the objective of the research reported in the present article is to obtain core-scale models (also known as Darcy-scale or macro-scale models) for permeable biofilm in two different pore geometries: a thin channel and a tube. The motivation for choosing these two geometries is because experiments are performed in the laboratory in micro-channels (\cite{Liu:Article:2019}) and tubes (\cite{Bott:Article:1983}), they may represent a fracture in a core sample (\cite{Bringedal:Article:2015}) and some porous media can be modelled as a stack of micro-tubes or micro-channels (\cite{Noorden:Article:2010}).

To summarize, the novel aspect in this work is the derivation of core-scale models from a pore-scale model for a biofilm which is permeable to the flow and has a variable (in time and space) height. The fluid flow in the biofilm is modelled by the Brinkman equation, whereas in the remaining pore space the Stokes model is adopted. This is done for two different geometries. We derive analytical expressions for the upscaled quantities and provide numerical simulations for the upscaled models in both cases. 

The structure of this paper is as follows. In Sec. 2, we describe the pore-scale biofilm model. In Sec. 3, we present the dimensionless pore-scale biofilm model. In Sec. 4, we perform formal homogenization on the model equations and obtain upscaled equations. In Sec. 5, we compare the upscaled models with the upscaled model of \cite{Noorden:Article:2010} and with the well-known core-scale model of \cite{Chen-Charpentier:Article:2009}. We compare the derived porosity-permeability relations to empirical porosity-permeability relations from the literature. Also, we perform numerical simulations in the upscaled models and we compare the results for the biofilm height and nutrient concentration for the two different effective models. Finally, in Sec. 6 we present the conclusions.

\section{Pore-scale model}\label{sec:pore-scale model}
The pore-scale mathematical model considered here follows ideas from \cite{Alpkvist:Article:2007} ,\cite{Noorden:Article:2010} and \cite{Deng:Article:2013}. A detailed description of this model can be found in \cite{Landa:Article:2019}, where a comparison of laboratory measurements and numerical simulations is also presented.

\begin{figure}
\includegraphics[width=\textwidth]{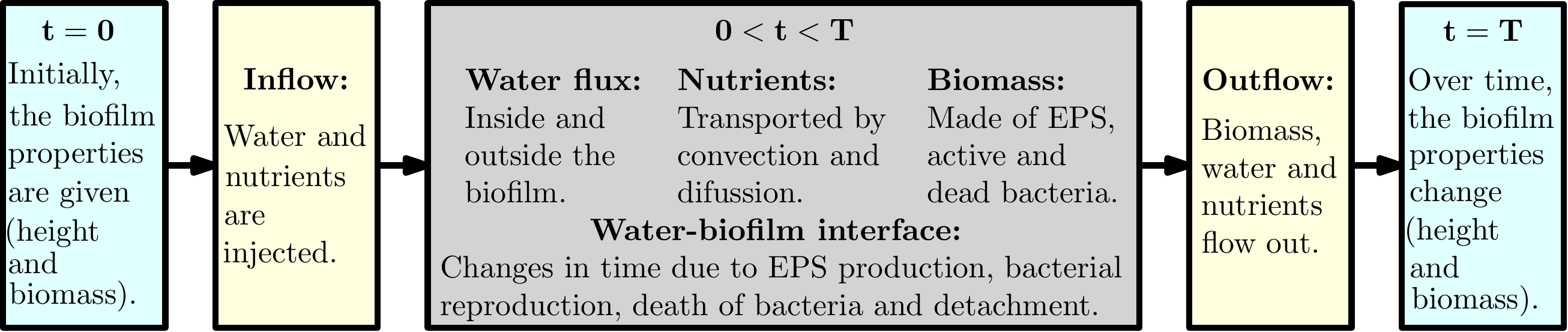}
\caption{Conceptual pore-scale model showing the processes for the biofilm dynamics.}
\label{fig:1}   
\end{figure}

The biofilm has four components: water, EPS, active and dead bacteria ($j=\lbrace w,e,a,d\rbrace$). Let $\theta_j$ and $\rho_j$ denote the volume fraction and the density of species $j$. The sum of volume fractions is constraint to 1 ($\theta_w+\theta_e+\theta_a+\theta_d=1$), where the volume fraction of water $\theta_w$ is assumed constant. The biomass phases and water are assumed to be incompressible ($\partial_t\rho_j=0$) and the biofilm layer is attached to the pore walls. Fig. \ref{fig:1} shows schematically the phenomena considered for the biofilm formation.

\begin{figure}
\includegraphics[width=\textwidth]{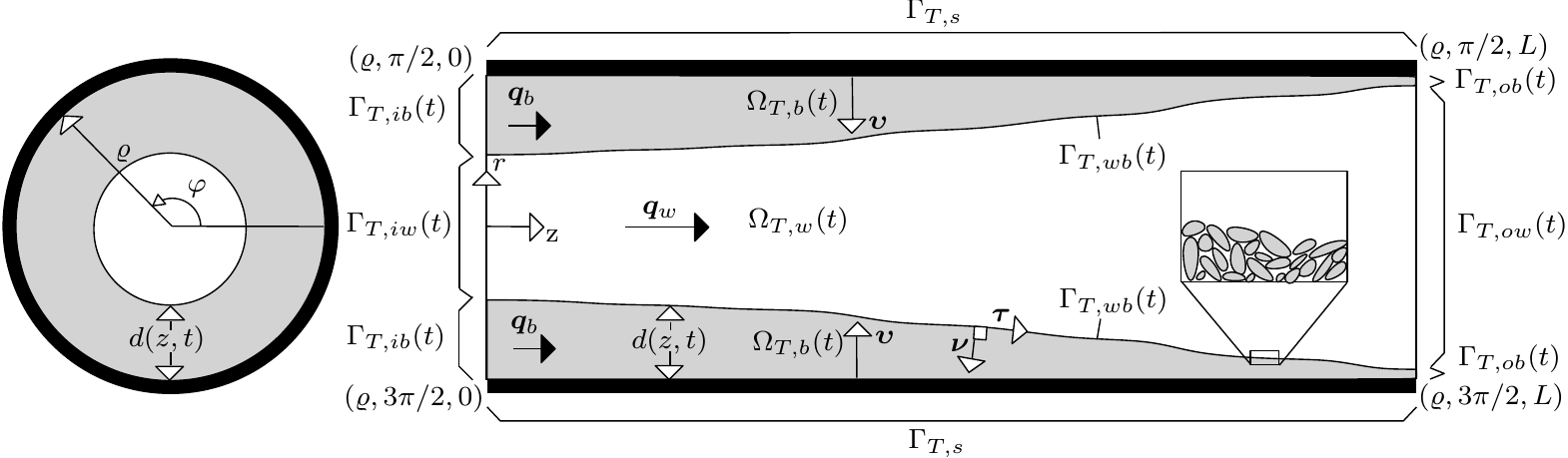}
\caption{Pore of radius $\varrho$ and length $L$ in cylindrical coordinates.}
\label{fig:2}   
\end{figure}

We consider two different pore geometries: a tube in cylindrical coordinates $\pmb{r}=(r,\varphi,z)$ and a thin channel in Cartesian coordinates $\pmb{x}=(x,y,z)$. The $z$ direction is taken along the length $L$ of the tube and thin channel (see Figs. \ref{fig:2} and \ref{fig:6}). In the first case, the pore has circular cross-section and in the second a rectangular one. In both cases, the length is much larger than the cross-sectional aperture. In both cases, we assume a certain symmetry. For the cylindrical pore we assume that the processes are radially symmetric, hence there is no angular dependence (see Fig. \ref{fig:2}). For the thin channel, there are no changes in the $x$ direction, i.e., the width of the channel, so it can be reduced to a two-dimensional strip (see Fig. \ref{fig:6}). This assumption is based on experiments, showing that when the width of the channel is much smaller than its height, the growth of the biofilm occurs only at the upper and lower walls along the channel  \cite{Liu:Article:2019}. We present in detail the upscaling of the model equations on the tube geometry. The upscaling on the channel geometry is shown in Appendix A. Fig. \ref{fig:2} shows the different domains, boundaries and interface in the pore with tubular geometry.
We consider a thin tube of radius $\varrho$ and length $L$. We denote the biofilm height by $d$ which only depends on $z$ and time as a result of the symmetry assumption. The domain is occupied by the water $\Omega_{T,w}(t)=\lbrace \pmb{r}|\;r\in[0,\varrho-d(z,t)),\;\varphi\in[0,2\pi),\;z\in(0,L) \rbrace$ and biofilm $\Omega_{T,b}(t)=\lbrace \pmb{r}|\;r\in(\varrho-d(z,t),\varrho),\;\varphi\in[0,2\pi),\;z\in(0,L) \rbrace$ phases with the biofilm located along the tube wall $\Gamma_{T,s}=\lbrace \pmb{r}|\;r=\varrho,\;\varphi\in[0,2\pi),\;z\in(0,L) \rbrace$. Clearly $r=\varrho-d(z,t)$ separates the water and biofilm regions. The water domain has three boundary parts: the inflow $\Gamma_{T,iw}(t) =\lbrace \pmb{r}|\;r\in [0,\varrho-d(z,t)),\;\varphi\in[0,2\pi),\;z=0\rbrace$, the outflow $\Gamma_{T,ow}(t) =\lbrace \pmb{r}|\;r\in [0,\varrho-d(z,t)),\;\varphi\in [0,2\pi),\;z=L\rbrace$ and the interface between the water and the biofilm $\Gamma_{T,wb}(t) =\lbrace \pmb{r}|\;r=\varrho-d(z,t),\;\varphi\in [0,2\pi),\;z\in(0,L)\rbrace$. Similarly, for the biofilm we have the inflow $\Gamma_{T,ib}(t)=\lbrace \pmb{r}|\;r\in (\varrho-d(z,t),\varrho),\;\varphi\in [0,2\pi),\;z=0\rbrace$ and outflow $\Gamma_{T,ob}(t)=\lbrace \pmb{r}|\;r\in (\varrho-d(z,t),\varrho),\;\varphi\in [0,2\pi)\;z=L\rbrace$ boundary parts, the water-biofilm interface $\Gamma_{T,wb}$ and the solid tube wall $\Gamma_{T,s}$. Although the tube is a three-dimensional domain, recalling the rotational symmetry, we only write the $r$- and $z$-components of the vectors in order to reduce the length of the mathematical expressions. 

The unit normal $\pmb{\nu}$ pointing into the biofilm and the normal velocity of the interface $\nu_n$ can be written in terms of the biofilm height $d$ as \cite{Noorden:Article:2010} 
\refstepcounter{equation}
$$
  \pmb{\nu}=(1,\partial_z d)/(1+(\partial_z d)^2)^{1/2},\quad\nu_n =-\partial_t d/(1+(\partial_z d)^2)^{1/2}.
  \eqno{(\theequation{\mathit{a},\mathit{b}})}\label{nuu}
$$
The water flux outside the biofilm $\Omega_{T,w}(t)$ is described by the Stokes and continuity equations
\refstepcounter{equation}
$$
\mu\Delta \pmb{q}_w=\nabla p_w,\quad \nabla\pmb{\cdot} \pmb{q}_w=0,
 \eqno{(\theequation{\mathit{a},\mathit{b}})}\label{waterp}
$$
while the water flux inside the biofilm $\Omega_{T,b}(t)$ is described by the Brinkman and continuity equations
\refstepcounter{equation}
$$
(\mu/\theta_w)\Delta \pmb{q}_b-(\mu/k)\pmb{q}_b=\nabla p_b,\quad\nabla\pmb{\cdot} \pmb{q}_b=0.
 \eqno{(\theequation{\mathit{a},\mathit{b}})}\label{waterpr}
$$
Here $p_w$ and $p_b$ are the water pressures and $\pmb{q}_w$ and $\pmb{q}_b$ are the water velocities in the water domain and biofilm domain respectively; $\mu$ is the water viscosity (constant, not dependent on biofilm species) and $k$ is the permeability of the biofilm (assumed isotropic). At the interface $\Gamma_{T,wb}(t)$ one has the continuity of the velocity and of the normal stress tensor
\refstepcounter{equation}
$$
\pmb{q}_{w}=\pmb{q}_{b},\quad\pmb{\nu}\pmb{\cdot} (\mu(\nabla\pmb{q}_{w}+\nabla\pmb{q}_{w}^T)-\mathbb{I}p_w)=\pmb{\nu}\pmb{\cdot} ((\mu/\theta_w)(\nabla\pmb{q}_{b}+\nabla\pmb{q}_{b}^T)-\mathbb{I}p_b), \eqno{(\theequation{\mathit{a},\mathit{b}})}\label{waterps}
$$
where $\mathbb{I}$ is the identity matrix. At the wall $\Gamma_{T,s}$ we consider the no-slip boundary condition $\pmb{q}_b=\pmb{0}$.

To model the nutrient transport and consumption, we let $c_\alpha$ $(\alpha\in\lbrace w,b\rbrace)$ stand for the nutrient concentration in water or biofilm (mass per total volume of biofilm) respectively and $D$ is the nutrient diffusion coefficient in water. Then, the nutrients in the water $\Omega_{T,w}(t)$ and the biofilm $\Omega_{T,b}(t)$ satisfy the convection-diffusion equation
\refstepcounter{equation}
$$
\partial_t c_{w}+\nabla\pmb{\cdot}\pmb{J}_w=0,\quad \partial_t(\theta_w c_{b})+\nabla\pmb{\cdot}\pmb{J}_b=R_{b},
 \eqno{(\theequation{\mathit{a},\mathit{b}})}\label{conp}
$$
where $\pmb{J}_w$ and $\pmb{J}_b$ are given by
\refstepcounter{equation}
$$
\pmb{J}_w=-D\nabla c_{w}+\pmb{q}_wc_{w},\quad \pmb{J}_b=-\theta_w D\nabla c_{b}+\pmb{q}_{b}c_{b}.
 \eqno{(\theequation{\mathit{a},\mathit{b}})}\label{jpp}
$$
Further at $\Gamma_{T,wb}(t)$ we impose the mass conservation and we assume continuity of nutrients
\refstepcounter{equation}
$$
(\pmb{J}_b-\pmb{J}_w)\pmb{\cdot} \pmb{\nu}=\nu_n(\theta_wc_b-c_w),\qquad \theta_wc_b=c_w.
 \eqno{(\theequation{\mathit{a},\mathit{b}})}\label{conpi}
$$
At the solid wall $\Gamma_{T,s}$ the normal flux is $\pmb{\upsilon}\pmb{\cdot}\pmb{J}_b=0$, where $\pmb{\upsilon}$ is the normal vector at the pore wall. The reaction term $R_b$ for the consumption of nutrients is given by
\begin{eqnarray}\label{rbp}
R_b=-\mu_n\theta_a\rho_ac_{b}/(k_n+c_{b}),
\end{eqnarray}
where $\mu_n$ is the maximum rate of nutrient consumption and $k_n$ is the Monod-half nutrient velocity coefficient.

The movement of the biomass components $\theta_i$ in $\Omega_{T,b}(t)$ due to reproduction, production of EPS and death of active bacteria can be modelled as a Darcy flow \cite{Alpkvist:Article:2007}. We denote by $\pmb{u}$ the velocity of the biomass and $\Phi$ the growth velocity potential. Then, we consider the following equations \cite{Alpkvist:Article:2007,Landa:Article:2019} 
\refstepcounter{equation}
$$
\pmb{u}=-\nabla\Phi,\quad \nabla\pmb{\cdot}\pmb{u}=(1-\theta_w)^{-1}\Sigma_i(R_i/\rho_i),\qquad i\in\lbrace e,a,d\rbrace
 \eqno{(\theequation{\mathit{a},\mathit{b}})}\label{gvp}
$$
The growth velocity potential is set to zero $\Phi=0$ at the interface $\Gamma_{T,wb}(t)$ and homogeneous Neumann boundary condition $\pmb{\upsilon}\pmb{\cdot}\nabla\Phi=0$ at the wall $\Gamma_{T,s}$. 

For each of the biomass components $\theta_i$ in $\Omega_{T,b}(t)$, we assume mass conservation \cite{Alpkvist:Article:2007}
\begin{eqnarray}\label{bcp}
\rho_i\partial_t\theta_i+\rho_i\nabla \pmb{\cdot} (\theta_i \pmb{u})=R_i,\qquad i\in\lbrace e,a,d\rbrace
\end{eqnarray}
and Neumann condition $\pmb{\nu}\pmb{\cdot}\nabla\theta_i=0$ at the interface $\Gamma_{T,wb}(t)$ and $\pmb{\upsilon}\pmb{\cdot}\nabla\theta_i=0$ at the wall $\Gamma_{T,s}$. The reaction terms for the biomass components are given by
\refstepcounter{equation}
$$
R_d=k_{res}\theta_a\rho_a,\quad R_e=-Y_eR_b,\quad R_a=-Y_aR_b-k_{res}\theta_a\rho_a,
 \eqno{(\theequation{\mathit{a},\mathit{b},\mathit{c}})}\label{sump}
$$
where $Y_e$ and $Y_a$ are yield coefficients and $k_{res}$ is the bacterial decay rate.\par

The water-biofilm interface changes in time due to changes inside the biofilm and the water flux provoking detachment of components, which is known as erosion. Thus, the normal velocity of the interface $\Gamma_{T,wb}(t)$ is given by \cite{Noorden:Article:2010}
\begin{equation}\label{nvp}
\nu_n=\left\{
\begin{array}{ll}
f^{+}(\pmb{\nu}\pmb{\cdot}\pmb{u}), &\qquad d=\varrho,\\
\pmb{\nu}\pmb{\cdot}\pmb{u}+k_{str}S, &\qquad 0<d<\varrho,\\
0, &\qquad d=0,
\end{array} \right.
\end{equation}
where we ensure that the biofilm-water interface does not overlap by taking the positive cut $f^{+}$ ($f^{+}(x):=\max(0,x)$) when $d=\varrho$. Here $k_{str}$ is the stress coefficient and $S$ is the tangential shear stress, given by \cite{Noorden:Article:2010}
\begin{eqnarray}\label{tsap}
S=||(\mathbb{I}-\pmb{\nu}{\pmb{\nu}}^T)\mu(\nabla \pmb{q}_w+\nabla \pmb{q}_w^T)\pmb{\nu}||.
\end{eqnarray}
This pore-scale model can be extended to consider more complex systems. For example, one can add different kind of nutrients, different active bacteria species in the biofilm or bacterial attachment.

\section{Non-dimensional model}\label{sec:non-dimensional model}
Before seeking an effective model, we bring the mathematical equations to a non-dimensional form. To this aim, we introduce the reference time $t_{ref}$, length $L_{ref}$, radius $\varrho_{ref}$, water velocity $q_{ref}=L_{ref}/t_{ref}$, biomass velocity $u_{ref}$, pressure $p_{ref}$ and concentration $c_{ref}$. The thin tube is characterized by the ratio of its radius to the length $\varepsilon=\varrho_{ref}/L_{ref}$, which is called the dimensionless aspect ratio. We define dimensionless coordinates and time as $\bar{r}=r/\varrho_{ref},\; \bar{z}=z/L_{ref}$ and $\bar{t}=t/t_{ref}$. The non-dimensional biofilm height is given by $\bar{d} =d/\varrho_{ref}$. The non-dimensional unit normal (\ref{nuu}) is given by
$\pmb{\bar{\nu}}(\bar{r},\bar{z})=(1,\varepsilon\partial_{\bar{z}} \bar{d})/(1+(\varepsilon\partial_{\bar{z}} \bar{d})^2)^{1/2}.$ We notice that a factor of $\varepsilon$ appears in the second component of the non-dimensional unit normal, as a result of the transformation of the coordinates
\[\partial_z d= \frac{1}{L_{ref}}\frac{\partial}{\partial \bar{z}}\bigg(\varrho_{ref} \frac{d}{\varrho_{ref}}\bigg)= \frac{1}{L_{ref}}\frac{\partial}{\partial \bar{z}}(\varrho_{ref} \bar{d})=\varepsilon\partial_{\bar{z}} \bar{d}.\]
Note that we have omitted the dependence of the vector variables on $\varphi$ ($\pmb{\bar{\nu}}(\bar{r},\varphi,\bar{z})=\pmb{\bar{\nu}}(\bar{r},\bar{z})$). This is justified by our assumption of the radial symmetry.  
The non-dimensional nutrient concentrations and densities are given by
$\bar{c}_w=c_w/c_{ref}$, $\bar{c}_b=c_b/c_{ref}$ and $\bar{\rho}_i=\rho_i/c_{ref}$ $(i\in \lbrace e,a,d\rbrace)$.

The water velocities are given by $\pmb{\bar{q}}_w(\bar{r},\bar{z})=(\bar{q}_{w,\bar{r}},\;\bar{q}_{w,\bar{z}}) =(q_{w,r}/(\varepsilon q_{ref}),q_{w,z}/q_{ref})$ and $\pmb{\bar{q}}_{b}(\bar{r},\bar{z})=(\bar{q}_{b,\bar{r}},\;\bar{q}_{b,\bar{z}}) =(q_{b,r}/(\varepsilon q_{ref}),q_{b,z}/q_{ref})$. The biomass velocity is given by
$\pmb{\bar{u}}(\bar{r},\bar{z}) =(\bar{u}_{\bar{r}},\;\bar{u}_{\bar{z}})=(u_r/(\varepsilon u_{ref}),u_z/u_{ref})$. We assume that the velocities in the radial direction are of the order $\rho_{ref}/t_{ref}$ (see \cite{Noorden:Article:2010}). Hence, they scale by $1/\varepsilon$ when compared to the longitudinal velocities. The biomass volume fractions are dimensionless; therefore, in the non-dimensional model we simply define $\bar{\theta}_i=\theta_i,\; i\in \lbrace w,e,a,d\rbrace$. Finally, the pressures and growth velocity potential become
$\bar{p}_w=p_w/p_{ref}$, $\bar{p}_{b} =p_{b}/p_{ref}$, $\bar{\Phi} =\Phi/(\varepsilon^2u_{ref}L_{ref})$.
We observe that the growth velocity potential $\Phi$ is scaled by $1/\varepsilon^2$ in order to have the biomass velocities in the radial direction of the order $\rho_{ref}/t_{ref}$ (see \cite{Noorden:Article:2010}). We define the following dimensionless parameters $\text{Pe}=q_{ref}L_{ref}/D$, $\bar{\mu}_n=t_{ref}\mu_{n}$, $\bar{k}_n=k_n/c_{ref}$, $\bar{k}=k/\varrho_{ref}^2$, $\bar{\mu}=\mu L_{ref} q_{ref}/(\varrho_{ref}^2 p_{ref})$, $\bar{k}_{str}=p_{ref}k_{str}/u_{ref}$ and $\bar{k}_{res}=t_{ref}k_{res}$.

In this way, the dimensionless system of equations for the water flux (\ref{waterp}-\ref{waterps}) is given by
\begin{eqnarray}
\frac{1}{\bar{r}}\partial_{\bar{r}}\left(\bar{r}\bar{q}_{w,\bar{r}}\right)+\partial_{\bar{z}}\bar{q}_{w,\bar{z}}&=&0,\label{T1}\\
\bar{\mu}\left(\frac{1}{\bar{r}}\partial_{\bar{r}}(\bar{r}\partial_{\bar{r}}\bar{q}_{w,\bar{r}})+\varepsilon^2\partial_{\bar{z}}^2 \bar{q}_{w,\bar{r}}-\frac{\bar{q}_{w,\bar{r}}}{\bar{r}^2}\right)&=&\varepsilon^{-2}\partial_{\bar{r}} \bar{p}_w,\label{T2}\\
\bar{\mu}\left(\frac{1}{\bar{r}}\partial_{\bar{r}}\left(\bar{r}\partial_{\bar{r}}\bar{q}_{w,\bar{z}}\right)+\varepsilon^2\partial_{\bar{z}}^2 \bar{q}_{w,\bar{z}}\right)&=&\partial_{\bar{z}} \bar{p}_w,\label{T3}\\
\frac{1}{\bar{r}}\partial_{\bar{r}}(\bar{r}\bar{q}_{b,\bar{r}})+\partial_{\bar{z}}\bar{q}_{b,\bar{z}}&=&0,\label{T4}\\
\frac{\bar{\mu}}{\bar{\theta}_w}\left(\frac{1}{\bar{r}}\partial_{\bar{r}}(\bar{r}\partial_{\bar{r}}\bar{q}_{b,\bar{r}})+\varepsilon^2\partial_{\bar{z}}^2 \bar{q}_{b,\bar{r}}-\frac{\bar{q}_{b,\bar{r}}}{\bar{r}^2}\right)&=&\frac{\bar{\mu}}{\bar{k}}\bar{q}_{b,\bar{r}}+\varepsilon^{-2}\partial_{\bar{r}} \bar{p}_b,\label{T5}\\
\frac{\bar{\mu}}{\bar{\theta}_w}\left(\frac{1}{\bar{r}}\partial_{\bar{r}}(\bar{r}\partial_{\bar{r}}\bar{q}_{b,{\bar{z}}})+\varepsilon^2\partial_{\bar{z}}^2 \bar{q}_{b,\bar{z}}\right)&=&\frac{\bar{\mu}}{\bar{k}}\bar{q}_{b,{\bar{z}}}+\partial_{\bar{z}} \bar{p}_b,\label{T6}\\
2\bar{\mu}\partial_{\bar{r}} \bar{q}_{w,\bar{r}}-\varepsilon^{-2}\bar{p}_w-2\frac{\bar{\mu}}{\bar{\theta}_w}\partial_{\bar{r}} \bar{q}_{b,\bar{r}}+\varepsilon^{-2}\bar{p}_b&=&\partial_{\bar{z}} \bar{d}\bigg(\frac{\bar{\mu}}{\bar{\theta}_w} \left(\partial_{\bar{r}}\bar{q}_{b,\bar{{z}}}+\varepsilon^2\partial_{\bar{z}}\bar{q}_{b,\bar{{r}}}\right)\nonumber\\
&\phantom{=}&-\bar{\mu}\left(\partial_{\bar{r}}\bar{q}_{w,\bar{z}}+\varepsilon^2\partial_{\bar{z}}\bar{q}_{w,\bar{r}}\right)\bigg),\label{T7}\\
\bar{\mu}\left(\partial_{\bar{r}}\bar{q}_{w,\bar{z}}+\varepsilon^2\partial_{\bar{z}}\bar{q}_{w,\bar{r}}\right)-\frac{\bar{\mu}}{\bar{\theta}_w} \left(\partial_{\bar{r}}\bar{q}_{b,\bar{{z}}}+\varepsilon^2\partial_{\bar{z}}\bar{q}_{b,\bar{{r}}}\right)&=&\partial_{\bar{z}} \bar{d}\bigg(2\varepsilon^{2}\frac{\bar{\mu}}{\bar{\theta}_w}\partial_{\bar{z}} \bar{q}_{b,\bar{z}}-\bar{p}_b\nonumber\\
&\phantom{=}&-2\varepsilon^{2}\bar{\mu}\partial_{\bar{z}} \bar{q}_{w,\bar{z}}+\bar{p}_w\bigg),\label{T8}\\
(\bar{q}_{w,\bar{r}},\;\bar{q}_{w,\bar{z}})&=&(\bar{q}_{b,\bar{r}},\;\bar{q}_{b,\bar{z}}),\label{T9}\\
(\bar{q}_{w,\bar{r}},\;\bar{q}_{w,\bar{z}})&=&(0,\;0),\label{T10}
\end{eqnarray} 
where (\ref{T1}-\ref{T3}) are the dimensionless Stokes and continuity equations, (\ref{T4}-\ref{T6}) are the dimensionless Brinkman and continuity equations, (\ref{T7}-\ref{T9}) are the dimensionless interface conditions and (\ref{T10}) is the dimensionless condition on the wall. 

The dimensionless equations for the transport of nutrients (\ref{conp}-\ref{conpi}) in the water and biofilm are given by
\begin{eqnarray}
\partial_{\bar{t}} \bar{c}_w-\frac{1}{\text{Pe}}\left(\frac{\varepsilon^{-2}}{{\bar{r}}}\partial_{\bar{r}}({\bar{r}}\partial_{\bar{r}} \bar{c}_w)+\partial_{\bar{z}}^2 \bar{c}_w\right)+\frac{1}{{\bar{r}}}\partial_{\bar{r}}(\bar{r}\bar{q}_{w,\bar{r}} \bar{c}_w)+\partial_{\bar{z}}(\bar{q}_{w,\bar{z}} \bar{c}_w)&=&0,\label{T11}\\
\partial_{\bar{t}} (\bar{\theta}_w \bar{c}_b)- \frac{\bar{\theta}_w}{\text{Pe}}\left(\frac{\varepsilon^{-2}}{{\bar{r}}}\partial_{\bar{r}}({\bar{r}}\partial_{\bar{r}} \bar{c}_b)+\partial_{\bar{z}}^2 \bar{c}_b\right)+\frac{1}{{\bar{r}}}\partial_{\bar{r}}(\bar{r}\bar{q}_{b,\bar{r}} \bar{c}_b)+\partial_{\bar{z}}(\bar{q}_{b,\bar{z}} \bar{c}_b)&=&\bar{R}_b,\label{T12}\\
-\frac{1}{\text{Pe}\varepsilon^{2}}(\partial_{\bar{r}} \bar{c}_w-\bar{\theta}_w \partial_{\bar{r}} \bar{c}_b)-(\bar{c}_b\bar{q}_{b,\bar{r}}-\bar{c}_w\bar{q}_{w,\bar{r}})+\partial_{\bar{t}} \bar{d}(\bar{\theta}_w\bar{c}_{b}-\bar{c}_w)\nonumber\\
+\frac{\partial_{\bar{z}} \bar{d}}{\text{Pe}}(\partial_{\bar{z}} \bar{c}_w-\bar{\theta}_w\partial_{\bar{z}} \bar{c}_b)+\partial_{\bar{z}} \bar{d}(\bar{c}_b\bar{q}_{b,\bar{z}}-\bar{c}_w\bar{q}_{w,\bar{z}})&=&0,\label{T13}\\
\bar{\theta}_w\bar{c}_b&=&\bar{c}_w,\label{T14}\\
\partial_r\bar{c}_b&=&0,\label{T14s}
\end{eqnarray}
where (\ref{T11}) is the dimensionless transport equation of nutrients in the water domain, (\ref{T12}) is the dimensionless transport equation of nutrients in the biofilm domain, (\ref{T13}-\ref{T14}) are the dimensionless coupling conditions at the interface and (\ref{T14s}) is the dimensionless condition on the wall. The dimensionless reaction rate (\ref{rbp}) for the consumption of nutrients is given by
$\bar{R}_b=-\bar{\mu}_n\bar{\theta}_a\bar{\rho}_a\bar{c}_b/(\bar{k}_n+\bar{c}_b)$.
The equations for the growth velocity potential (\ref{gvp}) become
\begin{eqnarray}
\frac{u_{ref}}{q_{ref}}\left(\frac{1}{{\bar{r}}}\partial_{\bar{r}}(\bar{r}\bar{u}_{\bar{r}})+\partial_{\bar{z}}\bar{u}_{\bar{z}}\right)&=&\bar{\Sigma},\label{T16}\\
(\bar{u}_{\bar{r}},\;\bar{u}_{\bar{z}})&=&-(\partial_{\bar{r}}\bar{\Phi},\varepsilon^2\partial_{\bar{z}}\bar{\Phi}),\label{T17}\\
\bar{\Phi}&=&0,\label{T18}\\
\partial_{\bar{r}}\bar{\Phi}&=&0,\label{T19}
\end{eqnarray} 
where (\ref{T16}-\ref{T17}) are the dimensionless equations for the biomass growth velocity potential, (\ref{T18}) is the dimensionless reference potential at the interface and (\ref{T19}) is the dimensionless condition on the wall. We define the dimensionless sum of the biomass reaction terms as $\bar{\Sigma}=(Y_e\bar{\rho}_a/\bar{\rho}_e+Y_a)\bar{\mu}_n\bar{\theta}_a\bar{c}_b/(\bar{k}_n+\bar{c}_b)+(\bar{\rho}_a/\bar{\rho}_{d}-1)\bar{k}_{res}\bar{\theta}_a.$
The equations for the biomass components (\ref{bcp}) become
\begin{eqnarray}
\partial_{\bar{t}}\bar{\theta}_e+\frac{u_{ref}}{q_{ref}}(\bar{u}_{\bar{r}}\partial_{\bar{r}}\bar{\theta}_e+\bar{u}_{\bar{z}}\partial_{\bar{z}}\bar{\theta}_e)&=&Y_e\bar{\mu}_n\bar{\theta}_a\frac{\bar{\rho}_a}{\bar{\rho}_{e}}\frac{\bar{c}_b}{\bar{k}_n+\bar{c}_b}-\bar{\theta}_e\bar{\Sigma},\label{T21}\\
\partial_{\bar{t}}\bar{\theta}_a+\frac{u_{ref}}{q_{ref}}(\bar{u}_{\bar{r}}\partial_{\bar{r}}\bar{\theta}_a+\bar{u}_{\bar{z}}\partial_{\bar{z}}\bar{\theta}_a)&=&Y_a\bar{\mu}_n\bar{\theta}_a\frac{\bar{c}_b}{\bar{k}_n+\bar{c}_b}-\bar{k}_{res}\bar{\theta}_a-\bar{\theta}_a\bar{\Sigma},\label{T22}\\
\partial_{\bar{t}}\bar{\theta}_d+\frac{u_{ref}}{q_{ref}}(\bar{u}_{\bar{r}}\partial_{\bar{r}}\bar{\theta}_d+\bar{u}_{\bar{z}}\partial_{\bar{z}}\bar{\theta}_d)&=&\bar{k}_{res}\frac{\bar{\rho}_a}{\bar{\rho}_{d}}\bar{\theta}_a-\bar{\theta}_d\bar{\Sigma},\label{T23}\\
-\partial_{\bar{r}}\bar{\theta}_i+\varepsilon\partial_{\bar{z}} \bar{d}\partial_{\bar{z}} \bar{\theta}_i&=&0\;\;\;i\in\lbrace e,a,d\rbrace,\label{T24}\\
\partial_{\bar{r}}\bar{\theta}_i&=&0\;\;\;i\in\lbrace e,a,d\rbrace,\label{T25}
\end{eqnarray}
where (\ref{T21}-\ref{T23}) are the dimensionless conservation of mass equations for the biomass components, (\ref{T24}) is the dimensionless condition at the interface and (\ref{T25}) is the dimensionless condition on the wall.

The dimensionless biofilm height (\ref{nvp}) is given by 
\begin{equation}\label{T26}
\partial_{\bar{t}} \bar{d}=\left\{\begin{array}{ll}
f^{+}((\bar{u}_{{\bar{r}}}-\partial_{\bar{z}}\bar{d} \bar{u}_{\bar{z}})u_{ref}/q_{ref}),&\qquad\bar{d}=1,\\
(\bar{u}_{{\bar{r}}}-\partial_{\bar{z}}\bar{d} \bar{u}_{\bar{z}})u_{ref}/q_{ref}-(1+(\varepsilon\partial_{\bar{z}} \bar{d})^2)^{1/2}\varepsilon \bar{k}_{str}\bar{S} ,&\qquad0<\bar{d}<1,\\
0,&\qquad\bar{d}=0.\end{array} \right.
\end{equation}
The dimensionless tangential shear stress (\ref{tsap}) is given by 
\begin{eqnarray}\label{T27}
\bar{S}&=||(\mathbb{I}-\pmb{\bar{\nu}}\pmb{\bar{\nu}}^{T})\bar{\mu} (\mathbb{\bar{M}}+\mathbb{\bar{M}}^T)\pmb{\bar{\nu}}||,
\end{eqnarray}
where the matrix $\mathbb{\bar{M}}$ is given by
\begin{eqnarray}
\mathbb{\bar{M}} &= 
\left( {\begin{array}{*{20}c}
\partial_{\bar{r}} \bar{q}_{w,\bar{r}} & \varepsilon\partial_{\bar{z}} \bar{q}_{w,\bar{r}}  \\
\varepsilon^{-1}\partial_{\bar{r}} \bar{q}_{w,\bar{z}} & \partial_{\bar{z}} \bar{q}_{w,\bar{z}}
\end{array} } \right).\label{T28}
\end{eqnarray}

\section{Upscaling}\label{sec:upscaling}
The pore-scale mathematical model describes the biofilm formation in a three-dimensional domain. Under some model assumptions, when the length of the tube is much larger than its radius, it is possible to reduce the dimensionality of the problem from three to one dimension, letting the aspect ratio $\varepsilon$ approach to zero. We perform a formal asymptotic expansion of the variables depending on $\varepsilon$, namely $\bar{p}_w$, $\bar{p}_b$, $\bar{c}_w$, $\bar{c}_b$, $\pmb{\bar{q}}_w$, $\pmb{\bar{q}}_b$, $\pmb{\bar{u}}$, $\bar{\Phi}$, $\bar{\theta}_w$, $\bar{\theta}_e$, $\bar{\theta}_a$, $\bar{\theta}_d$ and $\bar{d}$. For all except $\bar{d}$ we assume $\bar{\chi}(\pmb{{\bar{r}}},{\bar{t}})=\bar{\chi}_{0}(\pmb{{\bar{r}}},{\bar{t}})+\varepsilon \bar{\chi}_{1}(\pmb{{\bar{r}}},{\bar{t}})+O(\varepsilon^2)$. The corresponding asymptotic expansion of $d$ is
$\bar{d}({\bar{z}},{\bar{t}})=\bar{d}_{0}({\bar{z}},{\bar{t}})+\varepsilon \bar{d}_{1}({\bar{z}},{\bar{t}})+O(\varepsilon^2)$. In \cite{Noorden:Article:2010}, \cite{Kumar:Article:2014} and \cite{Bringedal:Article:2015}, the authors present upscaled models of pore-scale mathematical models for reactive flows. Following the same ideas, we can obtain the corresponding upscaled model in the tube pore geometry.

We define the average water velocity $\langle \bar{q} \rangle$ as the following integral
\begin{equation}\label{awf}
\langle\bar{q}\rangle({\bar{z}},{\bar{t}})=\langle\bar{q}_w\rangle({\bar{z}},{\bar{t}})+\langle\bar{q}_b\rangle({\bar{z}},{\bar{t}})=\frac{1}{\pi}\int_{0}^{2\pi}\bigg(\int_{0}^{1-\bar{d}_0}q_{w,\bar{z},0}\bar{r}d{\bar{r}}+\int_{1-\bar{d}_0}^{1}q_{b,\bar{z},0}\bar{r}d\bar{r}\bigg ) d\varphi.
\end{equation}
Notice that we divide by the cross-sectional area of the tube.
We consider the following space regions in the tube: 
\begin{eqnarray*}
\bar{\Xi}_w &=&\lbrace \pmb{\bar{r}}|\;\bar{r}\leq 1-\bar{d}\;\wedge\;0\leq\varphi<2\pi\;\wedge\;z_1\leq \bar{z} \leq z_1+\delta z\rbrace,\\
\bar{\Xi}_b &=&\lbrace \pmb{\bar{r}}|\;1-\bar{d}\leq \bar{r} \leq 1\;\wedge\;0\leq\varphi<2\pi\;\wedge\;z_1\leq \bar{z} \leq z_1+\delta z\rbrace. 
\end{eqnarray*}
These regions are a disk of radius $1-\bar{d}$ and a ring of thickness $\bar{d}$ respectively; both of length $\delta z$. Integrating (\ref{T1}) and (\ref{T4}) over the previous regions and using the Gauss's theorem, we obtain
\begin{eqnarray*}
0=&&\int_{\bar{\Xi}_w}\bar{\nabla}\pmb{\cdot} \pmb{\bar{q}}_w d\bar{V}+\int_{\bar{\Xi}_b}\bar{\nabla}\pmb{\cdot} \pmb{\bar{q}}_b d\bar{V}=2\pi\int_{z_1}^{z_1+\delta z}\pmb{\bar{q}}_{w}\pmb{\cdot}\pmb{\bar{\nu}}\big|_{{\bar{r}}=(1-\bar{d})}d\bar{z}\nonumber\\
&&+2\pi\int_{0}^{1-\bar{d}}(\bar{q}_{w,\bar{z}}\big|_{{\bar{z}}=z_1+\delta z}-\bar{q}_{w,\bar{z}}\big|_{{\bar{z}}=z_1})\bar{r}d\bar{r}-2\pi\int_{z_1}^{z_1+\delta z}(\pmb{\bar{q}}_{b}\pmb{\cdot}\pmb{\bar{\nu}}\big|_{{\bar{r}}=1-\bar{d})}+\pmb{\bar{q}}_{b}\pmb{\cdot}\pmb{\bar{\nu}}\big|_{{\bar{r}}=1})d\bar{z}\nonumber\\
&&+2\pi\int_{1-\bar{d}}^{1}(\bar{q}_{b,\bar{z}}\big|_{{\bar{z}}=z_1+\delta z}-\bar{q}_{b,\bar{z}}\big|_{\bar{z}=z_1})\bar{r}d\bar{r}.
\end{eqnarray*}
Recalling the no-slip condition  for the water flux on the wall (\ref{T10}) and the continuity of fluxes at the interface (\ref{T9}), the previous equation becomes
\begin{eqnarray*}
&&\int_{0}^{1-\bar{d}}(\bar{q}_{w,\bar{z}}\big|_{{\bar{z}}=z_1+\delta z}-\bar{q}_{w,\bar{z}}\big|_{{\bar{z}}=z_1})\bar{r}d\bar{r}+\int_{1-\bar{d}}^{1}(\bar{q}_{b,\bar{z}}\big|_{{\bar{z}}=z_1+\delta z}-\bar{q}_{b,\bar{z}}\big|_{{\bar{z}}=z_1})\bar{r}d\bar{r}=0.
\end{eqnarray*}
Dividing the previous equation by $\delta z$ and in the limit where $\delta z$ approach to zero, we obtain for the lowest-order terms in $\varepsilon$
\[\partial_{\bar{z}}\langle\bar{q}\rangle=\partial_{\bar{z}}\langle\bar{q}_w\rangle({\bar{z}},{\bar{t}})+\partial_{\bar{z}}\langle\bar{q}_b\rangle({\bar{z}},{\bar{t}})=0,\]
where we have used the definition of the average water velocity $\langle\bar{q}\rangle$ (\ref{awf}).

The lowest order terms in the Stokes model (\ref{T1}-\ref{T3}) lead to
\refstepcounter{equation}
$$
\partial_{\bar{r}}(\bar{r}\bar{q}_{w,\bar{r},0})/\bar{r}+\partial_{\bar{z}}\bar{q}_{w,\bar{z},0}=0,\quad\partial_{\bar{r}} \bar{p}_{w,0}=0,\quad\bar{\mu}\partial_{\bar{r}}({\bar{r}}\partial_{\bar{r}}\bar{q}_{w,\bar{z},0})/\bar{r}=\partial_{\bar{z}} \bar{p}_{w,0}.
\eqno{(\theequation{\mathit{a},\mathit{b},\mathit{c}})}\label{cyli}
$$
From (\ref{cyli}b), we conclude that $\bar{p}_{w,0}$ does not depend on the ${\bar{r}}$ coordinate. Analogously, for the Brinkman model (\ref{T4}-\ref{T6}), the lower-order terms in $\varepsilon$ give
\refstepcounter{equation}
$$
\partial_{\bar{r}}(\bar{r}\bar{q}_{b,\bar{r},0})/\bar{r}+\partial_{\bar{z}}\bar{q}_{b,\bar{z},0}=0,\quad\partial_{\bar{r}} \bar{p}_{b,0}=0,\quad\bar{\mu}\partial_{\bar{r}}({\bar{r}}\partial_{\bar{r}} \bar{q}_{b,\bar{z},0})/(\bar{r}\bar{\theta}_w)-\bar{\mu}\bar{q}_{b,\bar{z},0}/\bar{k}=\partial_{\bar{z}} \bar{p}_{b,0}.
\eqno{(\theequation{\mathit{a},\mathit{b},\mathit{c}})}\label{brinkc}
$$
From (\ref{brinkc}b), we conclude that $\bar{p}_{b,0}$ also does not depend on the ${\bar{r}}$ coordinate. Since neither $\bar{p}_{w,0}$ nor $\bar{p}_{b,0}$ depend on the ${\bar{r}}$ coordinate and from the lowest order terms in (\ref{T7}) we have that $\bar{p}_{w,0}=\bar{p}_{b,0}$ at the biofilm-water interface, we obtain that $\bar{p}_{w,0}({\bar{z}},{\bar{t}})=\bar{p}_{b,0}({\bar{z}},{\bar{t}})=\bar{p}_0({\bar{z}},{\bar{t}})$. We turn our attention to equations (\ref{cyli}c) and (\ref{brinkc}c). It is possible to find solutions for $q_{w,\bar{z},0}$ and $q_{b,\bar{z},0}$ integrating twice with respect to ${\bar{r}}$ both equations and in addition using the symmetry, interface and boundary conditions (\ref{T8}-\ref{T10}). After integration, we get
\refstepcounter{equation}
$$
\bar{q}_{w,\bar{z},0}=(\bar{r}^2/4+E )\partial_{\bar{z}}\bar{p}_0/\bar{\mu},\quad \bar{q}_{b,\bar{z},0}=(FJ_0(\xi {\bar{r}})+GY_0(-\xi {\bar{r}})-\bar{k} )\partial_{\bar{z}}\bar{p}_0/\bar{\mu},
\eqno{(\theequation{\mathit{a},\mathit{b}})}\label{bothb}
$$
where $J_\nu({\bar{z}})$ and $Y_\nu({\bar{z}})$ are the Bessel functions of order $\nu$ of first and second kind respectively (see \cite{Olver:Book:2012}). The coefficients appearing in (\ref{bothb}) are 
\begin{eqnarray*}
E=&&\frac{2w\bar{\theta}_w(J_0(\xi)Y_0(-\xi w)-J_0(\xi w)Y_0(-\xi))+\xi \bar{k}(J_0(\xi w)Y_1(-\xi w)+Y_0(-\xi w)J_1(\xi w))}{4(\xi J_0(\xi)Y_1(-\xi w)+\xi Y_0(-\xi)J_1(\xi w))}\\
&&-\frac{\xi(4\bar{k}+w^2)(J_0(\xi)Y_1(-\xi w)+Y_0(-\xi)J_1(\xi w))}{4(\xi J_0(\xi)Y_1(-\xi w)+\xi Y_0(-\xi)J_1(\xi w))},\\
F=&&\frac{2\bar{k}\xi Y_1(-\xi w)+w\bar{\theta}_wY_0(-\xi)}{2(\xi J_0(\xi)Y_1(-\xi w)+\xi Y_0(-\xi)J_1(\xi w))},\quad G=\frac{2\bar{k}\xi J_1(\xi w)+w\bar{\theta}_wJ_0(\xi)}{2(\xi J_0(\xi)Y_1(-\xi w)+\xi Y_0(-\xi)J_1(\xi w))},
\end{eqnarray*}
where $w=1-\bar{d}_0$, $\xi = i(\bar{\theta}_w/\bar{k})^{1/2}$ and $i$ is the imaginary number. We remark that most of the mathematical commercial software include Bessel functions; therefore, it is easy to use the above expression. Although the Bessel functions are evaluated with complex numbers, both fluxes $q_{w,{\bar{z}},0}$ and $q_{b,\bar{z},0}$ are real numbers.

To obtain the water velocity defined in (\ref{awf}), we integrate (\ref{bothb}) as follows 
\begin{eqnarray*}
\langle\bar{q}\rangle=&&\frac{\partial_{\bar{z}}\bar{p}_0}{\pi\bar{\mu}}\int_{0}^{2\pi}\bigg(\int_{0}^{1-\bar{d}_0}(\bar{r}^2/4+E )\bar{r}d\bar{r}\nonumber\\
&&+\int_{1-\bar{d}_0}^{1}(FJ_0(\bar{r}i(\bar{\theta}_w/\bar{k})^{1/2})+GY_0(-\bar{r}i(\bar{\theta}_w/\bar{k})^{1/2})-\bar{k})\bar{r}d\bar{r}\bigg )d\varphi\nonumber\\
=&&2 ((1-\bar{d}_0)^4/16+E(1-\bar{d}_0)^2/2+i(\bar{k}/\bar{\theta}_w)^{1/2}(FY_1(-i(\bar{\theta}_w/\bar{k})^{1/2})-GJ_1(i(\bar{\theta}_w/\bar{k})^{1/2})\nonumber\\
&&-F(1-\bar{d}_0)Y_1(-(1-\bar{d}_0)i(\bar{\theta}_w/\bar{k})^{1/2})+G(1-\bar{d}_0)J_1((1-\bar{d}_0)i(\bar{\theta}_w/\bar{k})^{1/2}))\nonumber\\
&&-\bar{k}(1-(1-\bar{d}_0)^2)/2 )\partial_{\bar{z}} \bar{p}_0/\bar{\mu}.
\end{eqnarray*}
This gives the Darcy's law $\langle\bar{q}\rangle=-\kappa_T(\bar{d}_0)\partial_{\bar{z}} p_0/\bar{\mu},$ where $\kappa_T(\bar{d}_0)$ is the effective permeability given by
\begin{eqnarray*}
\kappa_T(\bar{d}_0)=&&-2 ((1-\bar{d}_0)^4/16+E(1-\bar{d}_0)^2/2+i(\bar{k}/\bar{\theta}_w)^{1/2}(FY_1(-i(\bar{\theta}_w/\bar{k})^{1/2})\nonumber\\
&&-GJ_1(i(\bar{\theta}_w/\bar{k})^{1/2})-F(1-\bar{d}_0)Y_1(-(1-\bar{d}_0)i(\bar{\theta}_w/\bar{k})^{1/2})\nonumber\\
&&+G(1-\bar{d}_0)J_1((1-\bar{d}_0)i(\bar{\theta}_w/\bar{k})^{1/2}))-\bar{k}(1-(1-\bar{d}_0)^2)/2 ).
\end{eqnarray*}
which changes according to the biofilm height $\bar{d}_0$. 

The growth velocity potential equations (\ref{T16}) and (\ref{T17}) for the lower-order terms in $\varepsilon$ are
\refstepcounter{equation}
$$
u_{ref} [\partial_{\bar{r}}(\bar{r}\bar{u}_{\bar{r},0})/\bar{r}+\partial_{\bar{z}}\bar{u}_{\bar{z},0}]/q_{ref}=\bar{\Sigma}_0,\hspace{1cm} \bar{u}_{\bar{r},0}=-\partial_{\bar{r}}\bar{\Phi}_{b,0},\hspace{1cm} \bar{u}_{\bar{z},0}=0,
\eqno{(\theequation{\mathit{a},\mathit{b},\mathit{c}})}\label{velpottca}
$$
where the boundary conditions for the interface (\ref{T18}) becomes $\Phi_{b,0}=0$ and wall (\ref{T19}) becomes $\partial_{\bar{r}} \Phi_{b,0}=0$.

In dimensionless form, the volume fraction equations (\ref{T21}-\ref{T23}) are
\begin{equation}\label{mamat}
\partial_{\bar{t}}\bar{\theta}_i+u_{ref}(\bar{u}_{\bar{r}}\partial_{\bar{r}}\bar{\theta}_i+\bar{u}_{\bar{z}}\partial_{\bar{z}}\bar{\theta}_i)/q_{ref}=\bar{R}_i-\bar{\theta}_i\bar{\Sigma},\quad i=\lbrace e,a,d\rbrace.
\end{equation}
We focus on biofilms where the biomass components change slightly along the ${\bar{r}}$ direction, resulting in the approximation $\bar{\theta}_{i,0}({\bar{r}},{\bar{z}},{\bar{t}})=\bar{\theta}_{i,0}({\bar{z}},{\bar{t}})$. Using (\ref{velpottca}c), the lower-order terms in (\ref{mamat}) are
$\partial_{\bar{t}}\bar{\theta}_{i,0}=\bar{R}_{i,0}-\bar{\theta}_{i,0}\bar{\Sigma}_0$. Integrating (\ref{velpottca}a) over ${\bar{r}}$ and using the boundary conditions (\ref{T18}-\ref{T19}) one gets
\begin{equation}\label{sobrestc}
\bar{u}_{\bar{r},0}=q_{ref}\bar{\Sigma}_0({\bar{r}}-1)/2u_{ref}.
\end{equation}
For the nutrients, integrating (\ref{T11}) and (\ref{T12}) over ${\bar{r}}$ and $\varphi$ yields
\begin{eqnarray*}
2\pi\int_{0}^{1-\bar{d}}\bigg(\partial_{\bar{t}} \bar{c}_w-\frac{1}{\text{Pe}}\left(\varepsilon^{-2}\frac{1}{{\bar{r}}}\partial_{\bar{r}} ({\bar{r}}\partial_{\bar{r}} \bar{c}_w)+\partial_{\bar{z}}^2 \bar{c}_w\right)+\frac{1}{{\bar{r}}}\partial_{\bar{r}}(\bar{r}\bar{q}_{w,\bar{r}} \bar{c}_w)+\partial_{\bar{z}}(\bar{q}_{w,\bar{z}} \bar{c}_w)\bigg)\bar{r}d\bar{r}&&=0,\\
2\pi\int_{1-\bar{d}}^{1}\bigg(\partial_{\bar{t}} (\bar{\theta}_w \bar{c}_b)- \frac{\bar{\theta}_w}{\text{Pe}}\left(\varepsilon^{-2}\frac{1}{{\bar{r}}}\partial_{\bar{r}} ({\bar{r}}\partial_{\bar{r}} \bar{c}_b)+\partial_{\bar{z}}^2 \bar{c}_b\right)+\frac{1}{{\bar{r}}}\partial_{\bar{r}}(\bar{r}\bar{q}_{b\bar{r}} \bar{c}_b)+\partial_{\bar{z}}(\bar{q}_{b,\bar{z}} \bar{c}_b)\nonumber\\
+\bar{\mu}_n\bar{\theta}_a\bar{\rho}_a \frac{\bar{c}_b}{\bar{k}_n+\bar{c}_b}\bigg)\bar{r}d\bar{r}&&=0.
\end{eqnarray*}
Interchanging the integration and the differentiation operators, these equations become 
\begin{eqnarray*}
\partial_{\bar{t}}\bigg(\int_{0}^{1-\bar{d}} \bar{c}_w\bar{r}d\bar{r}\bigg)+\partial_{\bar{t}}\bar{d} \bar{r}\bar{c}_w\big|_{{\bar{r}}=1-\bar{d}}
-\partial_{\bar{z}}\bigg(\int_{0}^{1-\bar{d}}\bigg(\frac{1}{\text{Pe}}\partial_{\bar{z}} \bar{c}_w-\bar{q}_{w,\bar{z}}\bar{c}_w\bigg)\bar{r}d\bar{r}\bigg)\nonumber\\
-\partial_{\bar{z}}\bar{d}\bigg(\frac{1}{\text{Pe}}{\bar{r}}\partial_{\bar{z}} \bar{c}_w-\bar{r}\bar{q}_{w,\bar{z}}\bar{c}_w\bigg)\big|_{{\bar{r}}=1-\bar{d}}+\bigg(\frac{1}{\varepsilon^2\text{Pe}}{\bar{r}}\partial_{\bar{r}} \bar{c}_w-\bar{r}\bar{q}_{w,\bar{r}}\bar{c}_w \bigg)|_{{\bar{r}}=1-\bar{d}}&&=0,\label{nt11}\\
\partial_{\bar{t}}\bigg(\int_{1-\bar{d}}^{1}\bar{\theta}_w c_{b}^{\varepsilon}\bar{r}d\bar{r}\bigg)-\partial_{\bar{t}}\bar{d}\bar{\theta}_w \bar{c}_b{\bar{r}}\big|_{{\bar{r}}=1-\bar{d}}\nonumber\\
-\partial_{\bar{z}}\bigg(\int_{1-\bar{d}}^{1}\bigg(\frac{\bar{\theta}_w}{\text{Pe}}\partial_{\bar{z}} \bar{c}_b-\bar{q}_{b,\bar{z}}\bar{c}_b\bigg)\bar{r}d\bar{r}\bigg)+\partial_{\bar{z}}\bar{d}\bigg(\frac{\bar{\theta}_w}{\text{Pe}}{\bar{r}}\partial_{\bar{z}} \bar{c}_b-\bar{r}\bar{q}_{b,\bar{z}}\bar{c}_b\bigg)\big|_{{\bar{r}}=1-\bar{d}}\nonumber\\
-\bigg(\frac{\bar{\theta}_w}{\varepsilon^2\text{Pe}}{\bar{r}}\partial_{\bar{r}} \bar{c}_b-\bar{r}\bar{q}_{b,\bar{r}}\bar{c}_b \bigg)|_{{\bar{r}}=1-\bar{d}}+\bar{\mu}_n\bar{\rho}_a\bar{\theta}_a\int_{1-\bar{d}}^{1} \frac{\bar{c}_b}{\bar{k}_n+\bar{c}_b}\bar{r}d\bar{r}&&=0.\label{nt22}
\end{eqnarray*}
The lower order terms in the equations for the conservation of nutrients (\ref{T11}) and (\ref{T12}) are  $\partial_{\bar{r}}( {\bar{r}} \partial_{\bar{r}} \bar{c}_{w,0})=0$ and $\partial_{\bar{r}}({\bar{r}} \partial_{\bar{r}} (\bar{\theta}_w\bar{c}_{b,0}))=0$ respectively. The interface coupling condition (\ref{T13}) becomes $\partial_{\bar{r}}\bar{c}_{w,0}=\partial_{\bar{r}}(\bar{\theta}_w\bar{c}_{b,0})$ and (\ref{T14}) becomes $\bar{c}_{w,0}=\bar{\theta}_w\bar{c}_{b,0}$, while the boundary condition on the wall (\ref{T14s}) becomes $\partial_{\bar{r}}(\bar{\theta}_wc_{b,0})=0$. Therefore, we conclude that $\bar{c}_{w,0}({\bar{z}},{\bar{t}})=\bar{\theta}_w\bar{c}_{b,0}({\bar{z}},{\bar{t}})=\bar{c}_0({\bar{z}},{\bar{t}})$. Then, using the aforementioned results, the equations for the nutrients can be written as
\begin{eqnarray*}
\frac{1}{2}\partial_{\bar{t}}(\bar{c}_0(1-\bar{d}_0)^2)+\partial_{\bar{t}}\bar{d}_0 \bar{r}\bar{c}_0\big|_{{\bar{r}}=1-\bar{d}_0}-\frac{(1-\bar{d}_0)^2}{2\text{Pe}}\partial_{\bar{z}}^2 \bar{c}_0+\bar{c}_0\int_{0}^{1-\bar{d}_0}\bar{q}_{w,\bar{r},0}\bar{r}d\bar{r}\nonumber\\
-\partial_{\bar{z}}\bar{d}_0\bigg(\frac{1}{\text{Pe}}{\bar{r}}\partial_{\bar{z}} \bar{c}_0-\bar{r}\bar{q}_{w,\bar{z}}\bar{c}_0\bigg)\big|_{{\bar{r}}=1-\bar{d}_0}+\bigg(\frac{1}{\varepsilon^2\text{Pe}}{\bar{r}}\partial_{\bar{r}} \bar{c}_0-\bar{r}\bar{q}_{w,\bar{r},0}\bar{c}_0 \bigg)|_{{\bar{r}}=1-\bar{d}_0}&&=0,\\
\frac{1}{2}\partial_{\bar{t}}(\bar{c}_{0}(1-(1-\bar{d}_0)^2))-\partial_{\bar{t}}\bar{d}_0\bar{r} \bar{c}_0\big|_{{\bar{r}}=1-\bar{d}_0}-\frac{1-(1-\bar{d}_0)^2}{2\text{Pe}}\partial^2_{\bar{z}} \bar{c}_0\nonumber\\
+\bar{c}_0\int_{1-\bar{d}_0}^{1}\bar{q}_{b,\bar{z},0}\bar{r}d\bar{r}
+\partial_{\bar{z}}\bar{d}_0\bigg (\frac{1}{\text{Pe}}{\bar{r}}\partial_{\bar{z}} \bar{c}_0-\bar{r}\bar{q}_{b,\bar{z},0}\bar{c}_0\bigg)\big|_{{\bar{r}}=1-\bar{d}_0}\nonumber\\
-\bigg(\frac{1}{\varepsilon^2\text{Pe}}{\bar{r}}\partial_{\bar{r}} \bar{c}_0-\bar{q}_{b,\bar{r},0}\bar{c}_0\bigg)|_{{\bar{r}}=1-\bar{d}_0}+\frac{1-(1-\bar{d}_0)^2}{2}\bar{\mu}_n\bar{\rho}_a\bar{\theta}_{a,0}\frac{\bar{c}_0}{\bar{k}_n+\bar{c}_{0}}&&=0.
\end{eqnarray*}
Then, adding both previous equations and using the interface condition (\ref{T13}), we finally obtain
\begin{eqnarray*}
\partial_{\bar{t}}\bar{c}_0+\partial_{\bar{z}}(\bar{c}_0\langle \bar{q} \rangle-\partial_{\bar{z}} \bar{c}_0/\text{Pe})=-(1-(1-\bar{d}_0)^2)\bar{\mu}_n\bar{\theta}_{a,0}\bar{\rho}_a\bar{c}_0/(\bar{k}_n+\bar{c}_0).
\end{eqnarray*}
We focus on the water-biofilm interface (\ref{T26}):
\begin{equation}\label{macas}
\partial_{\bar{t}} \bar{d}=\left\{\begin{array}{ll}
f^{+}(u_{ref}(\bar{u}_{\bar{r}}-\partial_{\bar{z}}\bar{d} \bar{u}_{\bar{z}})/q_{ref}),&\qquad\bar{d}=1,\\
u_{ref}(\bar{u}_{\bar{r}}-\partial_{\bar{z}}\bar{d} \bar{u}_{\bar{z}})/q_{ref}-(1+(\varepsilon\partial_{\bar{z}} \bar{d})^2)^{1/2}\varepsilon \bar{k}_{str}\bar{S},&\qquad 0<\bar{d}<1,\\
0,&\qquad\bar{d}=0.\end{array}\right.
\end{equation}
Following \cite{Noorden:Article:2010}, we regularize the formulation of (\ref{macas}). First we let $H_{0}$ and $H_{1}$ be the set-valued Heaviside graphs
\begin{eqnarray}\label{hev}
H_{0}(\bar{d})=\left\{\begin{array}{ll}
\lbrace 0 \rbrace, &\qquad\bar{d}<0,\\
 \mathclose[0,1\mathclose] , &\qquad\bar{d}=0,\\
\lbrace 1 \rbrace, &\qquad\bar{d}>0,
\end{array}\right.
\qquad H_{1}(\bar{d})=\left\{\begin{array}{ll}
\lbrace 1 \rbrace, &\qquad\bar{d}<1,\\
\mathclose[0,1\mathclose], &\qquad\bar{d}=1,\\
\lbrace 0 \rbrace, &\qquad\bar{d}>1,
\end{array}\right.
\end{eqnarray}
where we set $H_0(\bar{d}=0)=0$ and $H_{1}(\bar{d}=1)=0$. Observe that this choice guarantees that $\partial_{\bar{t}} \bar{d}$ never becomes negative whenever $\bar{d}=0$ and positive when $\bar{d}=1$. Then, (\ref{macas}) is written as
\begin{eqnarray}\label{roberss}
\partial_{\bar{t}} \bar{d}\in && H_0(\bar{d})H_1(\bar{d}) ( u_{ref}(\bar{u}_{\bar{r}}-\partial_{\bar{z}}\bar{d} \bar{u}_{\bar{z}})/q_{ref}-(1+(\varepsilon\partial_{\bar{z}} \bar{d})^2)^{1/2}\varepsilon \bar{k}_{str}\bar{\mu} ||(\mathbb{I}-\pmb{\bar{\nu}}\pmb{\bar{\nu}}^{T})(\mathbb{\bar{M}}+\mathbb{\bar{M}}^T)\pmb{\bar{\nu}}|| )\nonumber\\
&&+(1-H_1(\bar{d}))f^{+}(u_{ref}(\bar{u}_{\bar{r}}-\partial_{\bar{z}}\bar{d} \bar{u}_{\bar{z}})/q_{ref}).
\end{eqnarray}
For practical calculations, the multi-valued functions are replaced by regularized Heaviside functions, defined by
\begin{equation}\label{hevr}
H_{\delta,0}(\bar{d})=\left\{\begin{array}{ll}
0, &\quad\bar{d}<0,\\
\bar{d}/\delta, &\quad\bar{d}\in [0,\delta],\\
1, &\quad\bar{d}>\delta,
\end{array}\right.\qquad
H_{\delta,1}(\bar{d})=\left\{\begin{array}{ll}
1, &\quad\bar{d}<1,\\
(1+\delta-\bar{d})/\delta, &\quad\bar{d}\in [1,1+\delta],\\
0, &\quad\bar{d}>1+\delta,
\end{array}\right.
\end{equation}
where $\delta$ is a small regularization parameter. Then, we can write (\ref{roberss}) as
\begin{eqnarray}\label{robers}
\partial_{\bar{t}} \bar{d}= && H_{\delta,0}(\bar{d})H_{\delta,1}(\bar{d}) ( u_{ref}(\bar{u}_{\bar{r}}-\partial_{\bar{z}}\bar{d} \bar{u}_{\bar{z}})/q_{ref}+(1-H_{\delta,1}(\bar{d}))f^{+}(u_{ref}(\bar{u}_{\bar{r}}-\partial_{\bar{z}}\bar{d}\bar{u}_{\bar{z}})/q_{ref})\nonumber\\
&&-(1+(\varepsilon\partial_{\bar{z}} \bar{d})^2)^{1/2}\varepsilon \bar{k}_{str}\bar{\mu} ||(\mathbb{I}-\pmb{\bar{\nu}}\pmb{\bar{\nu}}^{T})(\mathbb{\bar{M}}+\mathbb{\bar{M}}^T)\pmb{\bar{\nu}}||).
\end{eqnarray}
Using (\ref{T27}-\ref{T28}, \ref{velpottca}c, \ref{sobrestc}) for the lower-order terms we have
\begin{equation}\label{roberr}
\partial_{\bar{t}} \bar{d}= H_{\delta,0}(\bar{d})H_{\delta,1}(\bar{d})(\bar{d}_0\bar{\Sigma}_0/2)-\bar{k}_{str}\bar{\mu} |\partial_{\bar{r}}\bar{q}_{w,\bar{z},0}|+(1-H_{\delta,1}(\bar{d}))f^{+}(-\bar{\Sigma}_0/2).
\end{equation}
Using (\ref{bothb}a), we obtain
\[\partial_{\bar{t}} \bar{d}= H_{\delta,0}(\bar{d})H_{\delta,1}(\bar{d})(\bar{d}_0\bar{\Sigma}_0/2)-\bar{k}_{str}(1-\bar{d}_0)|\partial_{\bar{z}}\bar{p}_0|/2+(1-H_{\delta,1}(\bar{d}))f^{+}(-\bar{\Sigma}_0/2).\]
The original model is obtained when passing $\delta$ to zero  \cite{vanDuijn:Article:2004}, obtaining finally 
\[\partial_{\bar{t}} \bar{d}_0=
\left\{\begin{array}{ll}
f^{-}(\bar{\Sigma}_0/2), &\qquad\bar{d}_0=1,\\
-\bar{k}_{str}(1-\bar{d}_0)|\partial_{\bar{z}}{\bar{p}_0}|/2+\bar{d}_0\bar{\Sigma}_0/2, &\qquad 0<\bar{d}_0<1,\\
0, &\qquad\bar{d}_0=0,\\
\end{array}\right.\]
where $f^{-}$ is the negative cut ($f^{-}(x):=\min(0,x)$).
\section{Discussion and comparison with other biofilm models}
The extension from a channel (tube) to a  porous medium is done by considering a stack of channels (tubes) of void space and solid material \cite{Noorden:Article:2010}, where we denote by $\phi$ the porosity of the porous medium. As \cite{Steefel:Article:1994} and \cite{Noorden:Article:2010}, we assume that all tubes have the same diameter and are equally separated. Therefore, multiplying the upscaled model equations by $\phi$, the corresponding core-scale mathematical models are obtained. 
Table \ref{tab:1} shows the core-scale equations of the van Noorden model \cite{Noorden:Article:2010}, the porous medium formed by channels and the porous medium formed by tubes, where $v=\phi q$ is the Darcy velocity. The van Noorden model accounts for water flux, transport of nutrients and bacteria, bacterial attachment, detachment of biomass due to erosion, growth of biomass due to nutrient consumption and death of bacteria. In our model we do not include transport of bacteria and bacterial attachment. The reason is because the pore-scale model was built based on laboratory experiments, where only nutrients were injected after inoculation of bacteria \cite{Landa:Article:2019}. For the details of the upscaling on the thin channel domain, see Appendix A.
\begin{table}
\caption{Core-scale equations for the $^a$channel, $^b$tube and $^c$van Noorden models}
\label{tab:1}
\begin{center}
\begin{tabular}{lll}
\hline\noalign{\smallskip}
Name & Upscaled equation\\
\noalign{\smallskip}\hline\noalign{\smallskip}
Darcy$^a$& $v=-\phi\kappa_C (d)\partial_z p/\mu$,&$\partial_zv=0$\\
Darcy$^b$& $v=-\phi\kappa_T (d)\partial_z p/\mu$,&$\partial_zv=0$\\
Darcy$^c$& $v=-\phi(1-d)^3\partial_z p/3\mu$,&$\partial_zv=0$\\
Nutrients$^a$&$\partial_t(\phi c)+\partial_z(cv-\phi\partial_z c/\text{Pe})=-d\phi\rho_aR$\\
Nutrients$^b$&$\partial_t(\phi c)+\partial_z(cv-\phi\partial_z c/\text{Pe} )=-(1-(1-d)^2)\phi\rho_aR$\\
Nutrients$^c$& $\partial_t(\phi c)+\partial_z(cv-\phi\partial_z c/\text{Pe} )=-d\phi\rho_aR$\\
Height$^a$&$\partial_t d=
\left\{\begin{array}{ll}
f^{-}(\Sigma), &\hspace{1cm}d=1,\\
-k_{str}(1-d)|\partial_z{p}|+d\Sigma,&\hspace{1cm}0<d<1,\\
0, &\hspace{1cm}d=0.\\
\end{array}\right.$\\
Height$^b$&$\partial_t d=
\left\{\begin{array}{ll}
f^{-}(\Sigma/2), &\hspace{.35cm}d=1,\\
-k_{str}(1-d)|\partial_z{p}|/2+d\Sigma/2, &\hspace{.35cm}0<d<1,\\
0, &\hspace{.35cm}d=0.\\
\end{array}\right.$\\
Height$^c$& $\partial_t d=
\left\{\begin{array}{ll}
f^{-}(\Sigma), &\hspace{1cm}d=1,\\
-k_{str}(1-d)|\partial_z{p}|+d\Sigma, &\hspace{1cm}0<d<1,\\
0, &\hspace{1cm}d=0.\\
\end{array}\right.$\\
Bacteria$^{a,b}$&$\partial_t\theta_a=Y_a\mu_n\theta_ac/(k_n+c)-k_{res}\theta_a-\theta_a\Sigma$\\
EPS$^{a,b}$&$\partial_t\theta_e=(\rho_a/\rho_e)Y_e\mu_n\theta_ac(k_n+c)-\theta_e\Sigma$\\
Dead$^{a,b}$&$\partial_t\theta_d=\rho_ak_{res}\theta_a/\rho_d-\theta_d\Sigma$\\
Reactions$^{a,b}$ &$\Sigma=(Y_e\rho_a/\rho_e+Y_a)R+(\rho_a/\rho_{d}-1)k_{res}\theta_a$,&$ R=\mu_n\theta_ac/(k_n+c)$\\
Reactions$^c$ & $\Sigma=Y_aR-k_{res}$,&$ R=\mu_nc/(k_n+c)$\\
\noalign{\smallskip}\hline
\end{tabular}
\end{center}
\end{table}

From Table \ref{tab:1}, we observe that for the Darcy flow, the permeability is different for the three models. A discussion of these relations is given at the end of this Section. For the equations describing the transport of nutrients, the difference between the van Noorden and channel model is on the reaction term. For the van Noorden model the nutrient consumption depends on the biofilm height, while in the channel model depends also in the volume fraction of active bacteria. Comparing the tube and channel model, we observe a different function of the biofilm height, due to the cylindrical geometry. For the biofilm height, the difference between the van Noorden and channel model is on the $\Sigma$ term, where for the van Noorden model the total sum of reactions accounts for bacterial reproduction and decay, while in the channel structure it also accounts for EPS. Finally, for the bacterial, EPS and dead bacterial volume fractions, the model equations are the same for the channel and tube models, while for the van Noorden model the active bacterial volume fraction is constant with value 1. 

In \cite{Noorden:Article:2010}, the authors compared their upscaled model with a well-known macro-scale model by \cite{Taylor:Article:1990}, where a mathematical model for an impermeable single-species biofilm including flow, transport and reactions is built. We compare the two derived uspscaled models with a macro-scale model by \cite{Chen-Charpentier:Article:2009}, where a mathematical model for impermeable multi-species biofilm including water flow, transport of nutrients and reactions is built (Table \ref{tab:2}). In this model the authors considered a biofilm formed by two bacterial species which consume nutrients, reproduce and die. The porosity of the core decreases as the biofilm grows which decreases the permeability of the core. 
\begin{table}
 \caption{Core-scale equations for the model comparison \cite{Chen-Charpentier:Article:2009}}
  \label{tab:2}
\begin{center}
\begin{tabular}{ll}	
\hline\noalign{\smallskip}	
Name & Equation\\
\noalign{\smallskip}\hline\noalign{\smallskip}
Darcy& $v=-K\partial_z p,\hspace{.5cm} \partial_zv=0$\\
Nutrients& $\partial_tc+\partial_z(cv-D\partial_z c)=R$\\
Porosity& $\phi=\phi_O(1-\theta_B-\theta_K)$\\
Permeability& $K=K_O(\phi/\phi_O)^{\eta}$\\
Component B&$\partial_t\theta_B=Y_B\mu_B\theta_Bc/(k_B+c)-k_{res}\theta_B$\\
Component K&$\partial_t\theta_K=Y_K\mu_K\theta_Kc/(k_K+c)-k_{res}\theta_K$\\
Reactions & $R=-\mu_B\theta_Bc/(k_B+c)-\mu_K\theta_Kc/(k_K+c)$\\
\noalign{\smallskip}\hline
\end{tabular}
\end{center}
\end{table}

In Table \ref{tab:2}, $K$ is the permeability, $\theta_B$ and $\theta_K$ the volume fractions of the contaminant-degrading microbe and the strong biofilm-forming microbe, $\phi_O$ and $K_O$ the clean surface porosity and initial permeability respectively and $\eta$ an experimentally determined parameter. In \cite{Chen-Charpentier:Article:2009}, the porosity decreases as the component concentrations increases. In our models, the porosity in the porous medium decreases as the biofilm height increases. However, in \cite{Chen-Charpentier:Article:2009} the authors do not include the detachment effects. The permeability in both tube and channel models have different functions as a result of the different geometries and also because of the water flow inside the biofilm. Notice that there is a quartic function of the biofilm height in one of the permeability terms in the porous medium formed by tubes, as proposed in \cite{Suchomel:Article:1998} and \cite{Mostafa:Article:2007}. For the nutrient transport, those models, in addition to the \cite{Chen-Charpentier:Article:2009} model, have the same form with different reaction rates. 

\begin{figure}
\centerline{\includegraphics[scale=.6]{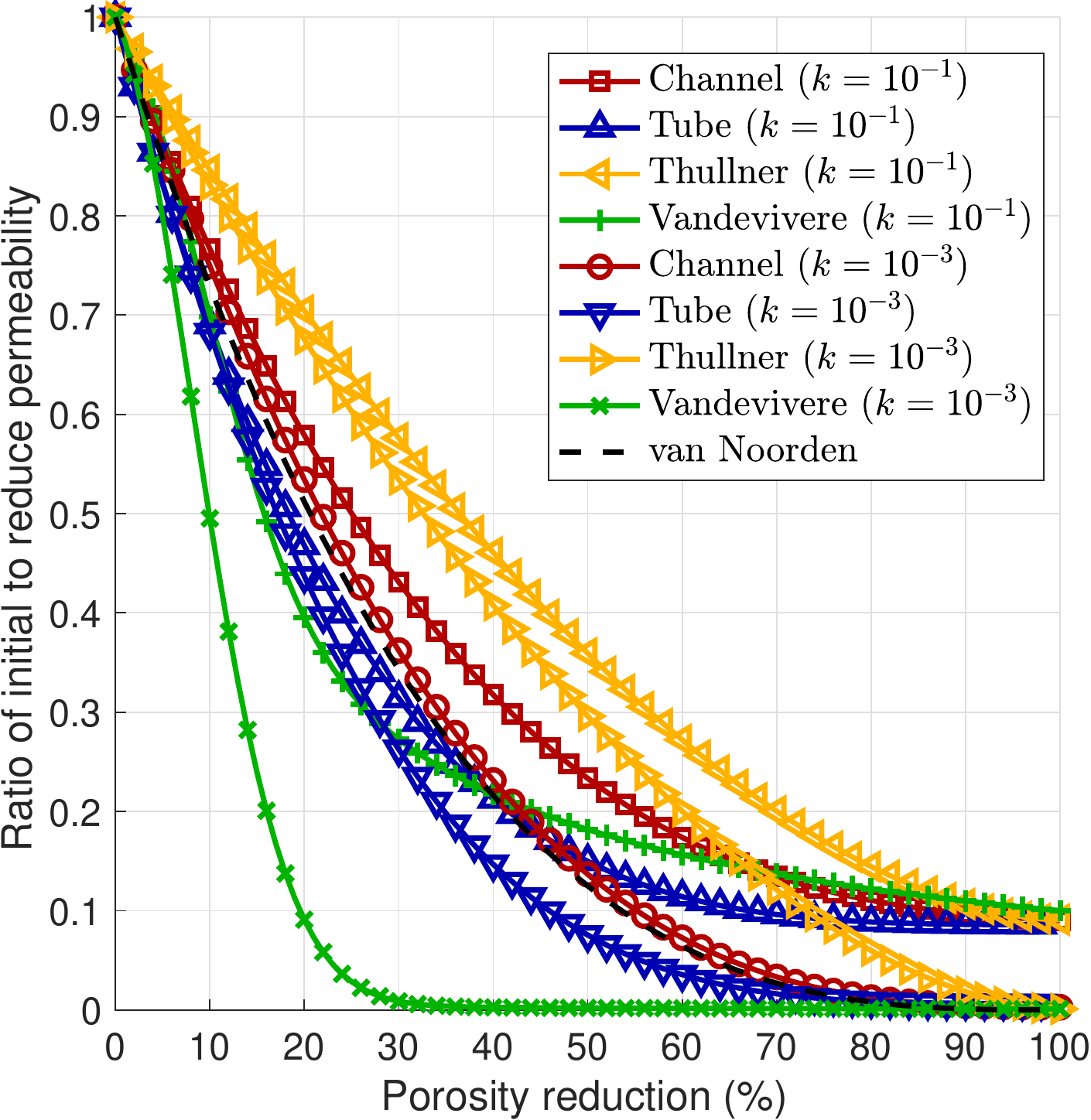}}
\caption{Ratio of initial to reduce permeability of different porosity-permeability relations for two different biofilm permeability values $k$.}
\label{fig:3}   
\end{figure}

Porosity-permeability relations for evolving pore space is an active research field (see \cite{Hommel:Article:2018} for a review of these relations). \cite{Thullner:Article:2002} present the following relation which includes the biofilm permeability $k$
\begin{equation}
K=K_O\bigg[\bigg(\frac{\phi-\phi_{crit}}{\phi_O-\phi_{crit}}\bigg)^\eta+k\bigg]\frac{1}{1+k},
\end{equation}
where $\phi_{crit}$ is the critical porosity at which the permeability becomes zero. \cite{Vandevivere:Article:1995} proposed the following relation of permeability and porosity for a plugging model
\begin{equation}
K=K_O[\exp(-0.5(B/B_c)^2)]\bigg(\frac{\phi}{\phi_O}\bigg)^2+[1-\exp(-0.5(B/B_c)^2)]\frac{k}{1-[(1-k/K_O)\phi/\phi_O]},
\end{equation}
where $B$ is a relative porosity given by $B=1-\phi/\phi_O$ and $B_c$ is the critical point where biofilm begins to detach and form plugs. Fig. \ref{fig:3} shows our derived porosity-permeability relations, the one derived by van Noorden and the two proposed relations by Thullner and Vandevivere for different values of biofilm permeability. The values of parameters are $\phi_{crit}=0$, $\eta=1.76$, $B_c=0.1$ \cite{Hommel:Article:2018} and $\theta_w=0.1$. For a biofilm with high permeability $k=10^{-1}$, we observe a faster reduction of permeability for Vandevivere. For a biofilm with low permeability $k=10^{-3}$, the channel and tube relations approach the one for van Noorden in contrast with the relations of Thullner and Vandevivere. In general, we observe different behaviours of the relations as the porosity decreases. 

We perform numerical simulations considering both effective models (channel and tube) to compare the biofilm height over time. We consider two different porous media of length $L=0.1$ m: the first one has pores formed by thin channels of height $2l=0.2$ mm and the second ones with tubes of diameter $2\varrho=0.2$ mm. For the inlet boundary, we set $p_i=4$ Pa. The injected nutrient concentration is $c_i=1$ kg m${}^{-3}$. The porosity $\Phi$ is set to 0.4. Recalling that biofilms are mostly composed by water, we set the water volume fraction in the biofilm equal to 90\%. We set the initial EPS and active bacterial volume fraction equal to 5\%; thus the initial dead bacterial volume fraction is 0. In Table \ref{tab:3}, the values of parameters for the numerical simulations are presented.
\begin{table}
\caption{Model parameters for the numerical studies}
\label{tab:3}
\begin{center}
\begin{tabular}{llll}	
\hline\noalign{\smallskip}	
Name & Description & Value & References\\
\noalign{\smallskip}\hline\noalign{\smallskip}
$\mu$	& Water viscosity &	$10^{-3}\;$Pa s & Well-known\\
$\rho_w$	& Water density &	$10^3\;$kg m${}^{-3}$ & Well-known\\
$\mu_n$ & Maximum growth rate&  $1.1\times 10^{-5}$ s$^{-1}$ & \cite{Alpkvist:Article:2007}\\
$k_n$	& Monod-half velocity&	$10^{-4}\;$kg m${}^{-3}$&\cite{Alpkvist:Article:2007}\\
$\rho_e$	& EPS density&	$60\;$kg m${}^{-3}$&\cite{Alpkvist:Article:2007}\\
$\rho_a$	& Bacterial density&	$60\;$kg m${}^{-3}$&\cite{Alpkvist:Article:2007}\\
$\rho_d$	& Dead bacterial density&	$60\;$kg m${}^{-3}$&\cite{Alpkvist:Article:2007}\\
$D$	& Nutrient diffusion&	$1.7\times 10^{-9}\;$m${}^2$ s$^{-1}$&\cite{Duddu:Article:2009}\\
$Y_{a}$ & Bacterial growth yield&	$0.553$&\cite{Duddu:Article:2009}\\
$Y_{e}$	& EPS growth yield&	$0.447$&\cite{Duddu:Article:2009}\\
$k_{res}$ & Bacterial decay rate&	$3.5\times 10^{-6}$ s$^{-1}$&\cite{Duddu:Article:2009}\\ 
$k$&Biofilm permeability&$10^{-9}\; $m${}^2$&\cite{Deng:Article:2013}\\
$k_{str}$&Stress coefficient&$2.6\times 10^{-10}$m (Pa s)$^{-1}$&\cite{Landa:Article:2019}\\
\noalign{\smallskip}\hline
\end{tabular}
\end{center}
\end{table}

We implement the model equations in the commercial software COMSOL Multiphysics (COMSOL 5.2a, Comsol Inc, Burlington, MA, www.comsol.com). A decoupled finite element algorithm is used to solve the mathematical model equations. Firstly, we solve for the pressure and concentration. Then, we compute the volume fractions and biofilm height. We iterate between both steps until the error (the difference between successive values of the solution) drops below a given tolerance $\zeta$. We perform numerical simulations and we compare the results of the two upscaled mathematical models.

\begin{figure}
\centerline{\includegraphics[scale=.5]{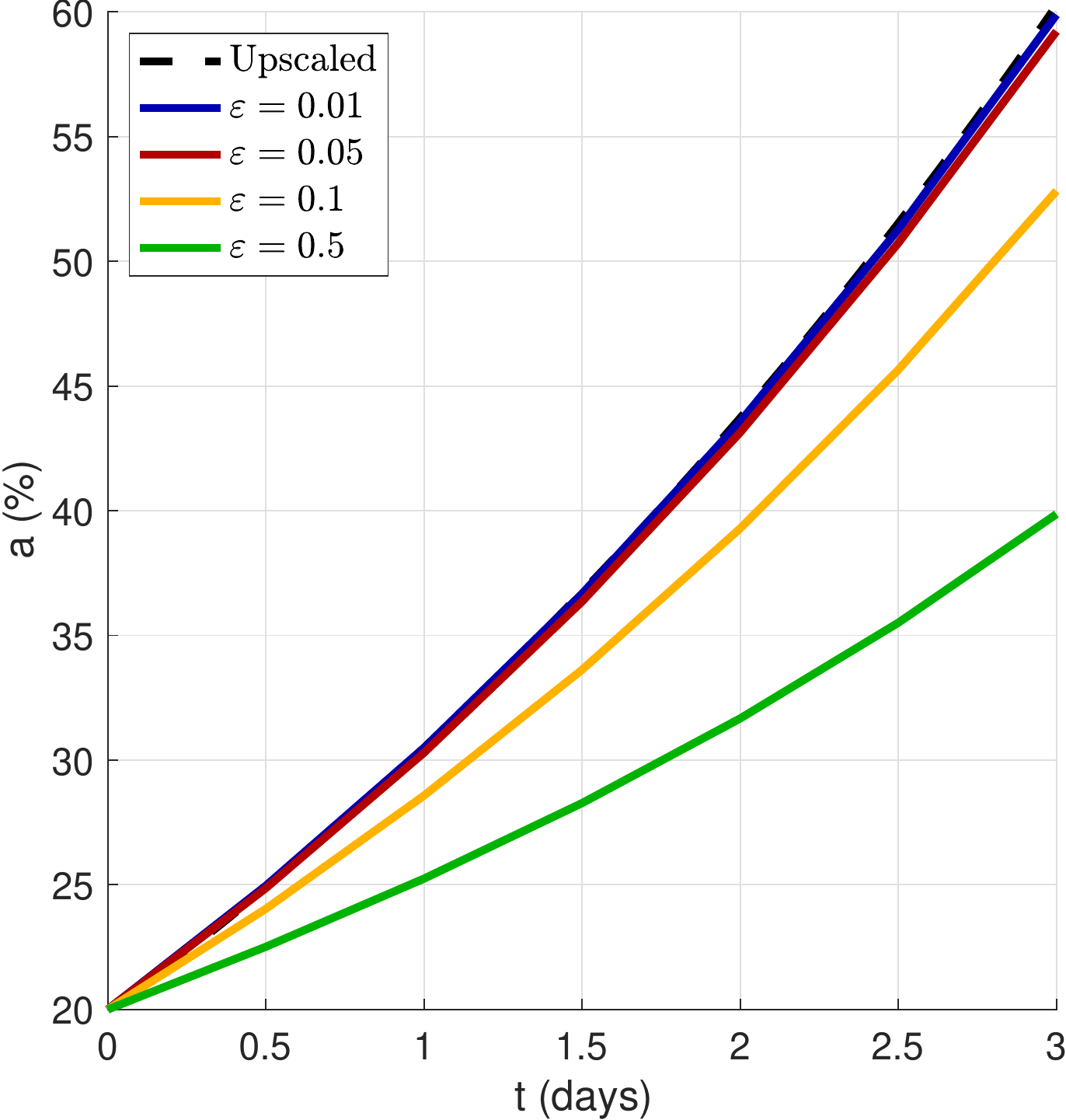}}
\caption{Percentage of biofilm coverage area over time for the upscaled model and for decreasing values of epsilon.}
\label{fig:4}
\end{figure}

To check if there is a correspondence between the pore-scale and upscaled models as $\varepsilon$ is close to zero, numerical simulations can be done for both models to compare the average solution of one of the variables. Fig. \ref{fig:4} compares the upscaled model with the pore-scale model in the channel for different values of $\varepsilon$, where the percentage of biofilm on the whole domain is plotted over time.  We called this coverage area $a$ and for the channel is given by 
\begin{equation}
a(t)=\int\limits_0^L d(z,t)dz/(lL).
\end{equation}
For all numerical simulations, we fixed the value of the height of the channel $2l$, where we set the initial biofilm height as $d=l/5$. Then, the length of the channel was changed accordingly to match the value of $\varepsilon$. We observe that the coverage area in the pore-scale simulations approaches the one computed from the upscaled models as $\varepsilon$ gets smaller. 

\begin{figure}
\centerline{\includegraphics[width=.5\textwidth]{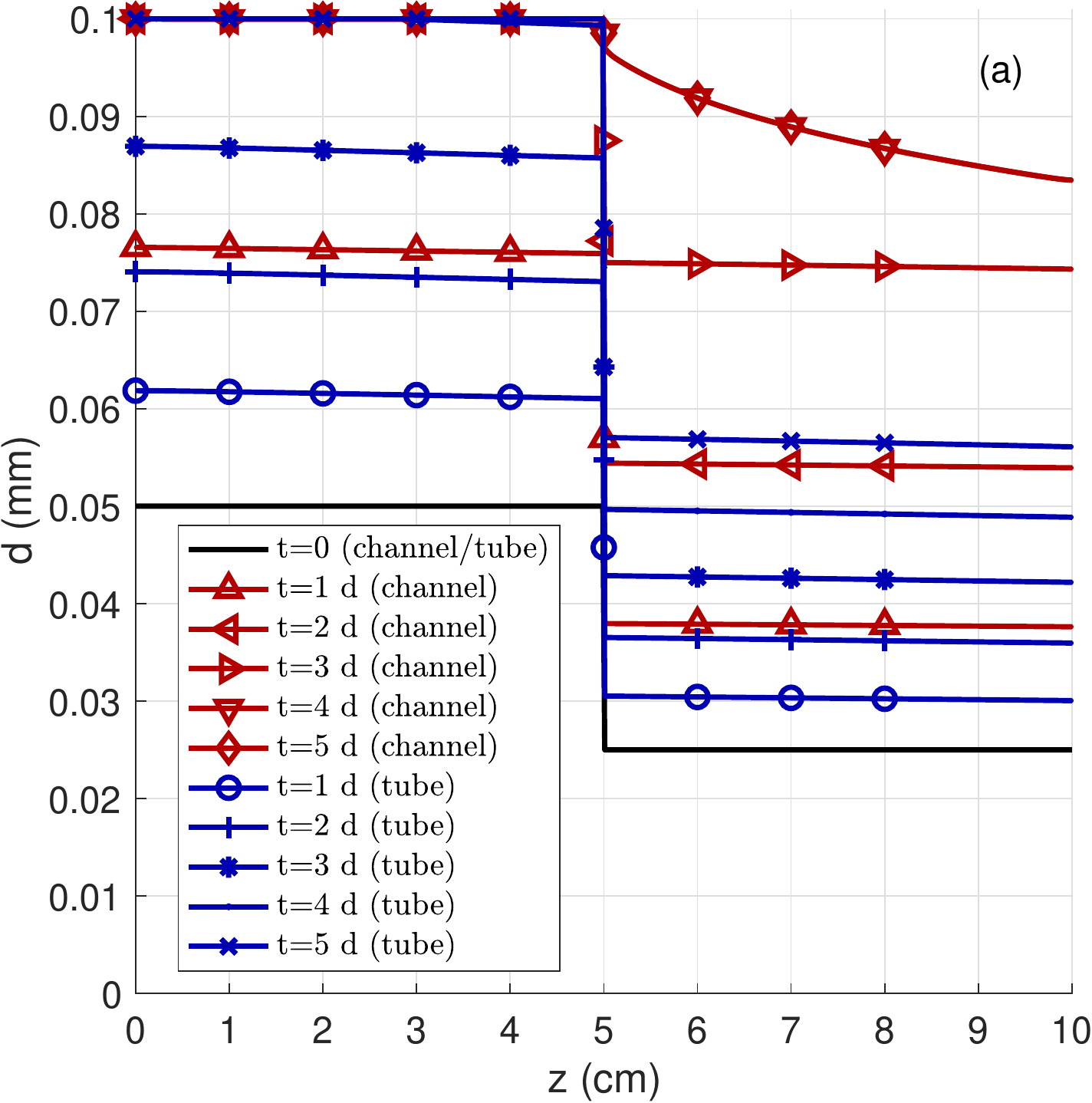}\includegraphics[width=.5\textwidth]{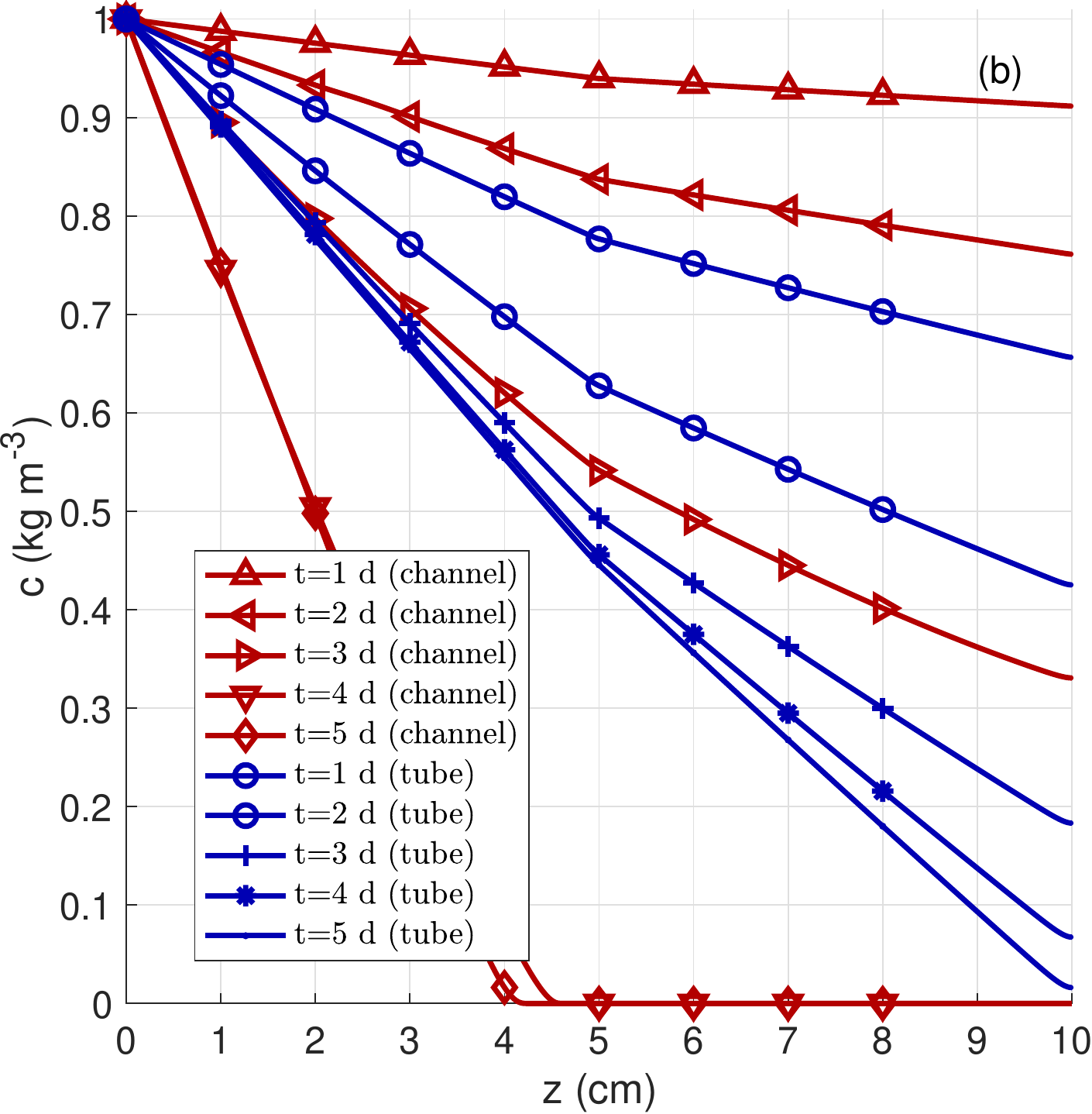}}
\caption{Biofilm heights (\textit{a}) and nutrient concentrations (\textit{b}) in the porous medium formed by channels and tubes.}
\label{fig:5}
\end{figure}

Fig. \ref{fig:5} shows the changes over time of biofilm height and nutrient concentration for both porous media. Initially, the left part ($0<z<L/2$) has a biofilm height of $d=l/2$ ($d=\varrho/2$ for the tubular pores), while the right part ($L>z>L/2$) has a height of $d=l/4$ ($d=\varrho/4$ for the tubular pores). This initial condition is given to study the biofilm development after clogging. We observe that the biofilm height increases faster for the pore channels than in the pore tubes. The explanation of this result is that the right-hand side of the equation for the biofilm height in the channel is two times greater than the right-hand side of the equation for the biofilm height in the tube which is obtained after upscaling in the two different geometries. For the porous medium formed by channels, we observe that the biofilm keeps growing even though the left part of the pore is clogged and after 4 days it reaches a stationary state. This result cannot be observed using the van Noorden model because the water flux stops once the channel is clogged. For the nutrients, we observe that the biofilm consumes the nutrients in the porous medium formed by tubes faster, but the biofilm in the porous medium formed by channels consumes more nutrients.  
\section{Conclusions}
In this work, we upscale a mathematical model for permeable biofilm considering a thin channel and tubular pore geometries. The upscaled models differ mainly in the effective permeability terms which are functions of the biofilm height. As $\varepsilon$ gets smaller, we obtain that the percentage of biofilm coverage area over time predicted by the pore-scale model approaches the one obtained
using the effective equations, which shows a correspondence between both models. After comparing with the model proposed by \cite{Noorden:Article:2010}, it is possible to derive this model as a particular case of the channel model. The derived upscaled models and the \cite{Chen-Charpentier:Article:2009} model are very similar. In this manner, the upscaling provides additional support for this model. The numerical simulations show that the biofilm height increases faster in the porous medium formed by channels than the one formed by tubes. These two upscaled models could be used to model porous media where the geometries of the fractures are similar to thin channels or tubes. To validate the core-scale upscaled models, designed laboratory experiments are necessary which is the subject of our future research.

\noindent\textbf{Acknowledgements} The work of D. Landa-Marb\'an, K. Kumar, G. B{\o}dtker and F. A. Radu was partially supported by the Research Council of Norway through the projects IMMENS no. 255426 and CHI no. 255510. I. S. Pop was supported by the Research Foundation-Flanders (FWO) through the Odysseus programme (project G0G1316N) and by Equinor through the Akademia grant. The authors would like to thank Brenna Connolly for improving the writing of the manuscript.

\appendix
\section{Upscaling of the mathematical model in a thin channel}\label{appA}
\begin{figure}
\centerline{\includegraphics[width=\textwidth]{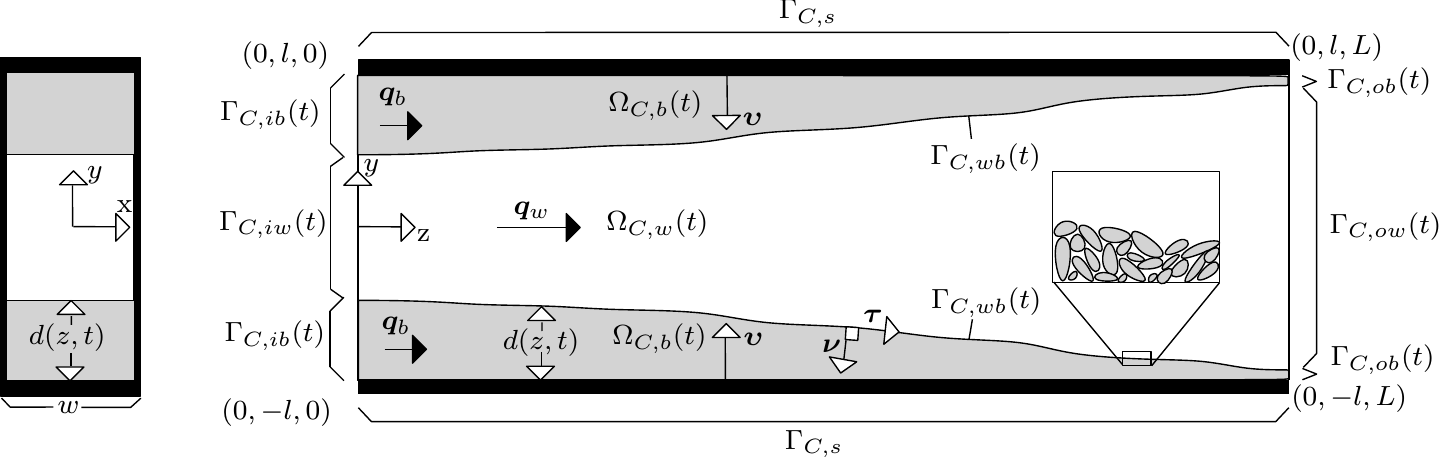}}
\caption{Pore of length $L$, height $2l$ and width $w$ in Cartesian coordinates.}
\label{fig:6}
\end{figure}

In Sec. 4, we show with details how to obtain the upscaled model equations in a tube. Following the same ideas, in this appendix we show how to upscale the model equations in a channel. We consider a thin channel with height $2l$, width $w$ and length $L$. When the width is much smaller than the height, experiments show that the growing of the biofilm occurs only in the upper and lower walls along the channel \cite{Liu:Article:2019}. Therefore, we can model the biofilm in the thin channel in a two-dimensional domain. Fig. \ref{fig:6} shows the different domains, boundaries and interface in the rectangular geometry.
To achieve non-dimensional quantities, we use the reference values defined in Sec. 4 ($t_{ref}$, $L_{ref}$, $q_{ref}$, $u_{ref}$, $p_{ref}$ and $c_{ref}$), where we consider the height of the channel $l_{ref}$ instead of the radius of the tube $\varrho_{ref}$. We define dimensionless coordinates and time as $\tilde{y}=y/l_{ref},\; \tilde{z}=z/L_{ref}$ and $\tilde{t}=t/t_{ref}$. The thin channel is characterized by the ratio of its height to the length $\epsilon=l_{ref}/L_{ref}$. All dimensionless variables and quantities are analogously defined as in Sec. 3, where we use $l_{ref}$ instead of $\varrho_{ref}$ and we denote the dimensionless variables with $\;\tilde{}\;$ instead of $\;\bar{}$.

The dimensionless system of equations for the water flux is given by
\begin{eqnarray}
\partial_{\tilde{y}}\tilde{q}_{w,\tilde{y}}+\partial_{\tilde{z}}\tilde{q}_{w,\tilde{z}}&=&0,\label{S1}\\
\tilde{\mu}(\epsilon^2\partial_{\tilde{z}}^2 \tilde{q}_{w,\tilde{y}}+\partial_{\tilde{y}}^2 \tilde{q}_{w,\tilde{y}})&=&\epsilon^{-2}\partial_{\tilde{y}} \tilde{p}_w,\label{S2}\\
\tilde{\mu}\epsilon^2\partial_{\tilde{z}}^2 \tilde{q}_{w,\tilde{z}}+\partial_{\tilde{y}}^2 \tilde{q}_{w,\tilde{z}}&=&\partial_{\tilde{z}} \tilde{p}_w,\label{S3}\\
\partial_{\tilde{y}}\tilde{q}_{b,\tilde{y}}+\partial_{\tilde{z}}\tilde{q}_{b,\tilde{z}}&=&0,\label{S4}\\
\frac{\tilde{\mu}}{\tilde{\theta}_w}(\epsilon^2\partial_{\tilde{z}}^2 \tilde{q}_{b,\tilde{y}}+\partial_{\tilde{y}}^2 \tilde{q}_{b,\tilde{y}})&=&\epsilon^{-2}\partial_{\tilde{y}} \tilde{p}_b+\frac{\tilde{\mu}}{\tilde{k}}\tilde{q}_{b,\tilde{y}},\label{S5}\\
\frac{\tilde{\mu}}{\tilde{\theta}_w}(\epsilon^2\partial_{\tilde{z}}^2 \tilde{q}_{b,\tilde{z}}+\partial_{\tilde{y}}^2 \tilde{q}_{b,\tilde{z}})&=&\partial_{\tilde{z}} \tilde{p}_b+\frac{\tilde{\mu}}{\tilde{k}}\tilde{q}_{b,\tilde{z}},\label{S6}\\
2\bar{\mu}\partial_{\bar{y}} \bar{q}_{w,\bar{y}}-\varepsilon^{-2}\bar{p}_w-2\frac{\bar{\mu}}{\bar{\theta}_w}\partial_{\bar{y}} \bar{q}_{b,\bar{y}}+\varepsilon^{-2}\bar{p}_b&=&\partial_{\bar{z}} \bar{d}\bigg(\frac{\bar{\mu}}{\bar{\theta}_w} \left(\partial_{\bar{y}}\bar{q}_{b,\bar{{z}}}+\varepsilon^2\partial_{\bar{z}}\bar{q}_{b,\bar{{y}}}\right)\nonumber\\
&\phantom{=}&-\bar{\mu}\left(\partial_{\bar{y}}\bar{q}_{w,\bar{z}}+\varepsilon^2\partial_{\bar{z}}\bar{q}_{w,\bar{y}}\right)\bigg),\label{S7}\\
\bar{\mu}\left(\partial_{\bar{y}}\bar{q}_{w,\bar{z}}+\varepsilon^2\partial_{\bar{z}}\bar{q}_{w,\bar{y}}\right)-\frac{\bar{\mu}}{\bar{\theta}_w} \left(\partial_{\bar{y}}\bar{q}_{b,\bar{{z}}}+\varepsilon^2\partial_{\bar{z}}\bar{q}_{b,\bar{{y}}}\right)&=&\partial_{\bar{z}} \bar{d}\bigg(2\varepsilon^{2}\frac{\bar{\mu}}{\bar{\theta}_w}\partial_{\bar{z}} \bar{q}_{b,\bar{z}}-\bar{p}_b\nonumber\\
&\phantom{=}&-2\varepsilon^{2}\bar{\mu}\partial_{\bar{z}} \bar{q}_{w,\bar{z}}+\bar{p}_w\bigg),\label{S8}\\
(\tilde{q}_{w,\tilde{y}},\;\tilde{q}_{w,\tilde{z}})&=&(\tilde{q}_{b,\tilde{y}},\;\tilde{q}_{b,\tilde{z}}),\label{S9}\\
(\tilde{q}_{b,\tilde{y}},\;\tilde{q}_{b,\tilde{z}})&=&(0,\;0).\label{S10}
\end{eqnarray}
The equations for the nutrients become
\begin{eqnarray}
\partial_{\tilde{t}} \tilde{c}_w-\frac{1}{\text{Pe}}(\epsilon^{-2}\partial_{\tilde{y}}^2 \tilde{c}_w+\partial_{\tilde{z}}^2 \tilde{c}_w)+\partial_{\tilde{y}}(\tilde{q}_{w,\tilde{y}} \tilde{c}_w)+\partial_{\tilde{z}}(\tilde{q}_{w,\tilde{z}} \tilde{c}_w)&=&0,\label{S11}\\
\partial_{\tilde{t}} (\tilde{\theta}_w \tilde{c}_b)- \frac{\tilde{\theta}_w}{\text{Pe}}(\epsilon^{-2}\partial_{\tilde{y}}^2 \tilde{c}_b+\partial_{\tilde{z}}^2 \tilde{c}_b)+\partial_{\tilde{y}}(\tilde{q}_{b,\tilde{y}} \tilde{c}_b)+\partial_{\tilde{z}}(\tilde{q}_{b,\tilde{z}} \tilde{c}_b)&=&\tilde{R}_b,\label{S12}\\
\hspace{-1cm}-\frac{1}{\text{Pe}\epsilon^{2}}(\partial_{\tilde{y}} \tilde{c}_w-\tilde{\theta}_w \partial_{\tilde{y}} \tilde{c}_b)-(\tilde{c}_b\tilde{q}_{b,\tilde{y}}-\tilde{c}_w\tilde{q}_{w,\tilde{y}})+\partial_{\tilde{t}} \tilde{d}(\tilde{\theta}_w\tilde{c}_{b}-\tilde{c}_w)\nonumber\\
+\frac{\partial_{\tilde{z}} \tilde{d}}{\text{Pe}}(\partial_{\tilde{z}} \tilde{c}_w-\tilde{\theta}_w\partial_{\tilde{z}} \tilde{c}_b)+\partial_{\tilde{z}} \tilde{d}(\tilde{c}_b\tilde{q}_{b,\tilde{z}}-\tilde{c}_w\tilde{q}_{w,\tilde{z}})&=&0,\label{S13}\\
\tilde{\theta}_w\tilde{c}_b&=&\tilde{c}_w,\label{S14}\\
\partial_{\tilde{y}}\tilde{c}_b&=&0,\label{S14s}
\end{eqnarray}
where $\tilde{R}_b=-\tilde{\mu}_n\tilde{\theta}_a\tilde{\rho}_a\tilde{c}_b/(\tilde{k}_n+\tilde{c}_b)$.
The dimensionless equations for the growth velocity potential are given by
\begin{eqnarray}
\frac{u_{ref}}{q_{ref}}(\partial_{\tilde{y}}\tilde{u}_{\tilde{y}}+\partial_{\tilde{z}}\tilde{u}_{\tilde{z}})&=&\tilde{\Sigma},\label{S16}\\
(\tilde{u}_{\tilde{y}},\;\tilde{u}_{\tilde{z}})&=&-(\partial_{\tilde{y}}\tilde{\Phi},\epsilon^2\partial_{\tilde{z}}\tilde{\Phi}),\label{S17}\\
\tilde{\Phi}&=&0,\label{S18}\\
\partial_{\tilde{y}}\tilde{\Phi}&=&0,\label{S19}
\end{eqnarray}
where $\tilde{\Sigma}=(Y_e\tilde{\rho}_a/\tilde{\rho}_e+Y_a)\tilde{\mu}_n\tilde{\theta}_a\tilde{c}_b/(\tilde{k}_n+\tilde{c}_b)+(\tilde{\rho}_a/\tilde{\rho}_{d}-1)\tilde{k}_{res}\tilde{\theta}_a$.
The equations for the biomass components become
\begin{eqnarray}
\hspace{-1cm}\partial_{\tilde{t}}\tilde{\theta}_e+\frac{u_{ref}}{q_{ref}}(\tilde{u}_{\tilde{y}}\partial_{\tilde{z}}\tilde{\theta}_e+\tilde{u}_{\tilde{z}}\partial_{\tilde{y}}\tilde{\theta}_e)&=&Y_e\tilde{\mu}_n\tilde{\theta}_a \frac{\tilde{\rho}_a}{\tilde{\rho}_{e}}\frac{\tilde{c}_b}{\tilde{k}_n+\tilde{c}_b}-\tilde{\theta}_e\tilde{\Sigma},\label{S21}\\
\hspace{-1cm}\partial_{\tilde{t}}\tilde{\theta}_a+\frac{u_{ref}}{q_{ref}}(\tilde{u}_{\tilde{y}}\partial_{\tilde{z}}\tilde{\theta}_a+\tilde{u}_{\tilde{z}}\partial_{\tilde{y}}\tilde{\theta}_a)&=&Y_a\tilde{\mu}_n\tilde{\theta}_a \frac{\tilde{c}_b}{\tilde{k}_n+\tilde{c}_b}-\tilde{k}_{res}\tilde{\theta}_a-\tilde{\theta}_a\tilde{\Sigma},\label{S22}\\
\hspace{-1cm}\partial_{\tilde{t}}\tilde{\theta}_d+\frac{u_{ref}}{q_{ref}}(\tilde{u}_{\tilde{y}}\partial_{\tilde{z}}\tilde{\theta}_d+\tilde{u}_{\tilde{z}}\partial_{\tilde{y}}\tilde{\theta}_d)&=&\tilde{k}_{res}\frac{\tilde{\rho}_a}{\tilde{\rho}_{d}}\tilde{\theta}_a-\tilde{\theta}_d\tilde{\Sigma},\label{S23}\\
-\partial_{\tilde{y}}\tilde{\theta}_i+\epsilon\partial_{\tilde{z}} \tilde{d}\partial_{\tilde{z}} \tilde{\theta}_i &=&0\;\;\;i\in\lbrace e,a,d\rbrace,\label{S24}\\
\partial_{\tilde{y}}\tilde{\theta}_i &=&0\;\;\;i\in\lbrace e,a,d\rbrace.\label{S25}
\end{eqnarray}
For the biofilm height we have
\begin{equation}
\partial_{\tilde{t}} \tilde{d}=\left\{\begin{array}{ll}
f^{+}(-\frac{u_{ref}}{q_{ref}}(-\tilde{u}_{\tilde{y}}+\partial_{\tilde{z}}\tilde{d} \tilde{u}_{\tilde{z}})),&\qquad\tilde{d}=1,\\
-(1+(\epsilon\partial_{\tilde{z}} \tilde{d})^2)^{1/2}\epsilon \tilde{k}_{str}\tilde{S}-\frac{u_{ref}}{q_{ref}}(-\tilde{u}_{\tilde{y}}+\partial_{\tilde{z}}\tilde{d} \tilde{u}_{\tilde{z}}),&\qquad 0<\tilde{d}<1,\\
0,&\qquad\tilde{d}=0,\end{array}\right.\label{S26}
\end{equation}
\raggedright
where
\begin{eqnarray}
\tilde{S}=||(\mathbb{I}-\pmb{\tilde{\nu}}\pmb{\tilde{\nu}}^{T})\tilde{\mu}( \mathbb{\tilde{M}}+\mathbb{\tilde{M}}^T)\pmb{\tilde{\nu}}^\epsilon||,\label{S27}
\end{eqnarray}
and 
\begin{eqnarray}
\mathbb{\tilde{M}}&= 
\left( {\begin{array}{*{20}c}
\partial_{\tilde{y}} \tilde{q}_{w,\tilde{y}} & \epsilon\partial_{\tilde{z}} \tilde{q}_{w,\tilde{y}}  \\
\epsilon^{-1}\partial_{\tilde{y}} \tilde{q}_{w,\tilde{z}} & \partial_{\tilde{z}} \tilde{q}_{w,\tilde{z}}
\end{array} } \right).\label{S28}
\end{eqnarray}
We define the average water velocity $\langle\tilde{q}\rangle$ as the following integral
\begin{equation}\label{awfc}
\langle\tilde{q}\rangle(\tilde{z},\tilde{t})=\langle\tilde{q}_w\rangle(\tilde{z},\tilde{t})+\langle\tilde{q}_b\rangle(\tilde{z},\tilde{t})=\frac{1}{2}\bigg (\int_{-(1-\tilde{d}_0)}^{1-\tilde{d}_0}\tilde{q}_{w,\tilde{z},0}d\tilde{y}+\int_{-1}^{-(1-\tilde{d}_0)}\tilde{q}_{b,\tilde{z},0}d\tilde{y}+\int_{1-\tilde{d}_0}^{1}\tilde{q}_{b,\tilde{z},0}d\tilde{y}\bigg).
\end{equation}
We define the following space regions in the channel
\begin{eqnarray*}
\tilde{\Xi}_u=&&\lbrace \pmb{\tilde{x}}|\;1-\tilde{d}\leq \tilde{y}\leq 1 \;\wedge\; z_1\leq \tilde{z} \leq z_1+\delta z\rbrace,\\
\tilde{\Xi}_m=&&\lbrace \pmb{\tilde{x}}|\;|\tilde{y}|\leq 1-\tilde{d}\;\wedge\; z_1\leq \tilde{z} \leq z_1+\delta z\rbrace,\\
\tilde{\Xi}_l=&&\lbrace \pmb{\tilde{x}}|\;-1\leq \tilde{y}\leq-(1-\tilde{d})\;\wedge\; z_1\leq \tilde{z} \leq z_1+\delta z\rbrace.
\end{eqnarray*}
Integrating (\ref{S1}) and (\ref{S4}) over the previous regions and using the Gauss's theorem yield
\begin{eqnarray*}
0=&&\int_{\tilde{\Xi}_u}\tilde{\nabla}\pmb{\cdot} \pmb{\tilde{q}}_b d\tilde{V}+\int_{\tilde{\Xi}_m}\tilde{\nabla}\pmb{\cdot} \pmb{\tilde{q}}_w d\tilde{V}+\int_{\tilde{\Xi}_l}\tilde{\nabla}\pmb{\cdot} \pmb{\tilde{q}}_b d\tilde{V}\nonumber\\
=&&2\int_{z_1}^{z_1+\delta z}\pmb{\tilde{q}}_{w}\pmb{\cdot}\pmb{\tilde{\nu}}\big|_{\tilde{y}=1-\tilde{d}}d\tilde{z}+\int_{-(1-\tilde{d})}^{1-\tilde{d}}(\tilde{q}_{w,\tilde{z}}\big|_{\tilde{z}=z_1+\delta z}-\tilde{q}_{w,\tilde{z}}\big|_{\tilde{z}=z_1})d\tilde{y}\nonumber\\
&&-\int_{z_1}^{z_1+\delta z}(\pmb{\tilde{q}}_b\pmb{\cdot}\pmb{\tilde{\nu}}\big|_{\tilde{y}=-(1-\tilde{d})}+\pmb{\tilde{q}}_b\pmb{\cdot}\pmb{\tilde{\nu}}\big|_{\tilde{y}=-1})d\tilde{z}+\int_{-1}^{-(1-\tilde{d})}(\tilde{q}_{b,\tilde{z}}\big|_{\tilde{z}=z_1+\delta z}-\tilde{q}_{b,\tilde{z}}\big|_{\tilde{z}=z_1})d\tilde{y}\nonumber\\
&&-\int_{z_1}^{z_1+\delta z}(\pmb{\tilde{q}}_b\pmb{\cdot}\pmb{\tilde{\nu}}\big|_{\tilde{y}=1}+\pmb{\tilde{q}}_b\pmb{\cdot}\pmb{\tilde{\nu}}\big|_{\tilde{y}=1-\tilde{d}})d\tilde{z}+\int_{1-\tilde{d}}^{1}(\tilde{q}_{b,\tilde{z}}\big|_{\tilde{z}=z_1+\delta z}-\tilde{q}_{b,\tilde{z}}\big|_{\tilde{z}=z_1})d\tilde{y}.
\end{eqnarray*}
Recalling the no-slip condition for the water flux on the wall (\ref{S10}) and the continuity of fluxes at the interface (\ref{S9}), the previous equation becomes
\begin{eqnarray*}
\int_{-(1-\tilde{d})}^{1-\tilde{d}}\left(\tilde{q}_{w,\tilde{z}}\big|_{\tilde{z}=z_1+\delta z}-\tilde{q}_{w,\tilde{z}}\big|_{\tilde{z}=z_1}\right)d\tilde{y}+\int_{-1}^{-(1-\tilde{d})}\left(\tilde{q}_{b,\tilde{z}}\big|_{\tilde{z}=z_1+\delta z}-\tilde{q}_{b,\tilde{z}}\big|_{\tilde{z}=z_1}\right)d\tilde{y}\nonumber\\
+\int_{1-\tilde{d}}^{1}\left(\tilde{q}_{b,\tilde{z}}\big|_{\tilde{z}=z_1+\delta z}-\tilde{q}_{b,\tilde{z}}\big|_{\tilde{z}=z_1}\right)d\tilde{y}&&=0.
\end{eqnarray*}
Dividing the previous equation by $\delta z$ and letting $\delta z$ approach zero, we obtain for the lowest-order terms in $\epsilon$
\[\partial_{\tilde{z}}\langle\tilde{q}\rangle=\partial_{\tilde{z}}\langle\tilde{q}_w\rangle(\tilde{z},\tilde{t})+\partial_{\tilde{z}}\langle\tilde{q}_b\rangle(\tilde{z},\tilde{t})=0,\]
where we have used the definition of the water velocity $\langle\tilde{q}\rangle$ (\ref{awfc}). 

The lowest order terms in the Stokes model (\ref{S1}-\ref{S3}) leads to
\refstepcounter{equation}
$$
\partial_{\tilde{y}}\tilde{q}_{w,\tilde{y},0}+\partial_{\tilde{z}}\tilde{q}_{w,\tilde{z},0}=0,\quad\partial_{\tilde{y}} \tilde{p}_{w,0}=0,\quad\tilde{\mu}\partial^2_{\tilde{y}}\tilde{q}_{w,\tilde{z},0}=\partial_{\tilde{z}} \tilde{p}_{w,0}.
\eqno{(\theequation{\mathit{a},\mathit{b},\mathit{c}})}\label{stokc}
$$
From (\ref{stokc}b), we conclude that $\tilde{p}_{w,0}$ does not depend on the $\tilde{y}$ coordinate. Analogously, for the Brinkman model (\ref{S4}-\ref{S6}), the lower-order terms in $\epsilon$ give
\refstepcounter{equation}
$$
\partial_{\tilde{y}}\tilde{q}_{b,\tilde{y},0}+\partial_{\tilde{z}}\tilde{q}_{b,\tilde{z},0}=0,\quad\partial_{\tilde{y}} \tilde{p}_{b,0}=0,\quad\tilde{\mu}\partial^2_y\tilde{q}_{b,\tilde{z},0}/\tilde{\theta}_w-\tilde{\mu}\tilde{q}_{b,\tilde{z},0}/\tilde{k}=\partial_{\tilde{z}} \tilde{p}_{b,0}.
\eqno{(\theequation{\mathit{a},\mathit{b},\mathit{c}})}\label{brinkcc}
$$
From (\ref{brinkcc}b), we obtain that $\tilde{p}_{b,0}$ does not depend on the $\tilde{y}$ coordinate and from the lowest order terms in (\ref{S7}) we conclude that $\tilde{p}_{w,0}(\tilde{z},\tilde{t})=\tilde{p}_{b,0}(\tilde{z},\tilde{t})=\tilde{p}_0(\tilde{z},\tilde{t})$. Integrating twice (\ref{stokc}) and (\ref{brinkcc}) with respect to $\tilde{y}$ and using the symmetry, interface and boundary conditions (\ref{S8}-\ref{S10})
\refstepcounter{equation}
$$
\tilde{q}_{w,\tilde{z},0}=\bigg (\frac{\tilde{y}^2}{2}+V\bigg )\frac{\partial_{\tilde{z}}\tilde{p}_0}{\tilde{\mu}},\quad\tilde{q}_{b,\tilde{z},0}=\bigg (We^{\tilde{y}(\tilde{\theta}_w/\tilde{k})^{1/2}}+Xe^{-\tilde{y}(\tilde{\theta}_w/\tilde{k})^{1/2}}-\tilde{k}\bigg)\frac{\partial_{\tilde{z}}\tilde{p}_0}{\tilde{\mu}},
\eqno{(\theequation{\mathit{a},\mathit{b}})}\label{chacab}
$$
where the coefficients are given by 
\begin{eqnarray*}
V=&&-\frac{(\frac{h^2}{2}+\tilde{k})(e^{-\tilde{d}_0(\tilde{\theta}_w/\tilde{k})^{1/2}}+e^{\tilde{d}_0(\tilde{\theta}_w/\tilde{k})^{1/2}})+(\tilde{k}\tilde{\theta}_w)^{1/2}h
    (e^{-\tilde{d}_0(\tilde{\theta}_w/\tilde{k})^{1/2}}-e^{\tilde{d}_0(\tilde{\theta}_w/\tilde{k})^{1/2}})-2\tilde{k}}{e^{-\tilde{d}_0(\tilde{\theta}_w/\tilde{k})^{1/2}}+e^{\tilde{d}_0(\tilde{\theta}_w/\tilde{k})^{1/2}}},\\
W=&&\frac{\tilde{k}e^{-h(\tilde{\theta}_w/\tilde{k})^{1/2})}+(\tilde{k}\tilde{\theta}_w)^{1/2}he^{(\tilde{\theta}_w/\tilde{k})^{1/2}}}{e^{-\tilde{d}_0(\tilde{\theta}_w/\tilde{k})^{1/2}}+e^{\tilde{d}_0(\tilde{\theta}_w/\tilde{k})^{1/2}}},\quad X=\frac{\tilde{k}e^{h(\tilde{\theta}_w/\tilde{k})^{1/2})}-(\tilde{k}\tilde{\theta}_w)^{1/2}he^{-(\tilde{\theta}_w/\tilde{k})^{1/2}}}{e^{-\tilde{d}_0(\tilde{\theta}_w/\tilde{k})^{1/2}}+e^{\tilde{d}_0(\tilde{\theta}_w/\tilde{k})^{1/2}}},\\
\end{eqnarray*}
where $h=-1+\tilde{d}_0$.

To obtain the water velocity defined in (\ref{awfc}), we integrate (\ref{chacab}) as follows
\begin{eqnarray*}
\langle\tilde{q}\rangle=&&\frac{\partial_{\tilde{z}}\tilde{p}_0}{2\tilde{\mu}}\bigg(\int_{-(1-\tilde{d}_0)}^{1-\tilde{d}_0}\bigg(\frac{\tilde{y}^2}{2}+V\bigg )d\tilde{y}+2\int_{-1}^{-(1-\tilde{d}_0)}\bigg(We^{\tilde{y}(\tilde{\theta}_w/\tilde{k})^{1/2}}+Xe^{-\tilde{y}(\tilde{\theta}_w/\tilde{k})^{1/2}}-\tilde{k}\bigg)d\tilde{y}\bigg )\nonumber\\
=&& ((1-\tilde{d}_0)^3/6+V(1-\tilde{d}_0)+((\tilde{k}/\tilde{\theta}_w)^{1/2}We^{-(\tilde{\theta}_w/\tilde{k})^{1/2}}(e^{\tilde{d}_0(\tilde{\theta}_w/\tilde{k})^{1/2}}-1)\nonumber\\
&&-(\tilde{k}/\tilde{\theta}_w)^{1/2}Xe^{(\tilde{\theta}_w/\tilde{k})^{1/2}}(e^{-\tilde{d}_0(\tilde{\theta}_w/\tilde{k})^{1/2}}-1)-\tilde{k}\tilde{d}_0) )\partial_{\tilde{z}} \tilde{p}_0/\tilde{\mu}\nonumber\\
=&&-\kappa_C(\tilde{d}_0)\partial_{\tilde{z}} \tilde{p}_0/\tilde{\mu}.
\end{eqnarray*}
This is the Darcy's law $\langle\tilde{q}\rangle=-\kappa_C(\tilde{d}_0)\partial_{\tilde{z}} \tilde{p}_0/\tilde{\mu}$, where $\kappa_C(\tilde{d}_0)$ is the effective permeability given by
\begin{eqnarray*}
\kappa_C(\tilde{d}_0)=&&- ((1-\tilde{d}_0)^3/6+V(1-\tilde{d}_0)+((\tilde{k}/\tilde{\theta}_w)^{1/2}We^{-(\tilde{\theta}_w/\tilde{k})^{1/2}}(e^{\tilde{d}_0(\tilde{\theta}_w/\tilde{k})^{1/2}}-1)\nonumber\\
&&-(\tilde{k}/\tilde{\theta}_w)^{1/2}Xe^{(\tilde{\theta}_w/\tilde{k})^{1/2}}(e^{-\tilde{d}_0(\tilde{\theta}_w/\tilde{k})^{1/2}}-1)-\tilde{k}\tilde{d}_0)).
\end{eqnarray*}

The growth velocity potential equations (\ref{S16}) and (\ref{S17}) for the lower-order terms in $\epsilon$ are
\begin{equation}\label{velpot}
u_{ref}(\partial_{\tilde{y}}\tilde{u}_{\tilde{y},0}+\partial_{\tilde{z}}\tilde{u}_{\tilde{z},0})/q_{ref}=\tilde{\Sigma}_0,\hspace{1cm} \tilde{u}_{\tilde{y},0}=-\partial_{\tilde{y}}\tilde{\Phi}_0,\hspace{1cm} \tilde{u}_{\tilde{z},0}=0,
\end{equation}
where the conditions at the interface (\ref{S18}) becomes $\tilde{\Phi}_0=0$ and wall (\ref{S19}) becomes $\partial_{\tilde{y}} \tilde{\Phi}_{0}=0$.

In dimensionless form, the volume fraction equations (\ref{S21}-\ref{S23}) are
\begin{equation}\label{mama}
\partial_{\tilde{t}}\tilde{\theta}_i+u_{ref}(\tilde{u}_{\tilde{y}}\partial_{\tilde{y}}\tilde{\theta}_i+\tilde{u}_{\tilde{z}}\partial_{\tilde{z}}\tilde{\theta}_i)/q_{ref}=\tilde{R}_i-\tilde{\theta}_i\tilde{\Sigma},
\end{equation}
with $i=\lbrace e,a,d\rbrace$. We focus on biofilms where the biomass components change slightly along the $\tilde{y}$ direction, resulting in the approximation $\tilde{\theta}_{i,0}(\tilde{y},\tilde{z},\tilde{t})=\tilde{\theta}_{i,0}(\tilde{z},\tilde{t})$. Using (\ref{velpot}c), the lower-order terms in (\ref{mama}) are
\begin{equation}\label{massa}
\partial_{\tilde{t}}\tilde{\theta}_{i,0}=\tilde{R}_{i,0}-\tilde{\theta}_{i,0}\tilde{\Sigma}_0.
\end{equation}
Integrating (\ref{velpot}a) over $\tilde{y}$ and using the boundary conditions (\ref{S18}-\ref{S19}) one gets
\begin{equation}\label{sobres}
\tilde{u}_{\tilde{y},0}=q_{ref}\tilde{\Sigma}_0(\tilde{y}+1)/u_{ref}.
\end{equation}
For the nutrients, integrating (\ref{S11}) and (\ref{S12}) yields
\begin{eqnarray*}
\int_{-(1-\tilde{d})}^{1-\tilde{d}}(\partial_{\tilde{t}} \tilde{c}_w-(\epsilon^{-2}\partial_{\tilde{y}}^2 \tilde{c}_w+\partial_{\tilde{z}}^2 \tilde{c}_w)/\text{Pe}+\partial_{\tilde{y}}(\tilde{q}_{w,\tilde{y}} \tilde{c}_w)+\partial_{\tilde{z}}(\tilde{q}_{w,\tilde{z}} \tilde{c}_w))d\tilde{y}&&=0,\\
2\int_{-1}^{-(1-\tilde{d})}(\partial_{\tilde{t}} (\tilde{\theta}_w \tilde{c}_b)- \tilde{\theta}_w(\epsilon^{-2}\partial_{\tilde{y}}^2 \tilde{c}_b+\partial_{\tilde{z}}^2 \tilde{c}_b)/\text{Pe}+\partial_{\tilde{y}}(\tilde{q}_{b,\tilde{y}} \tilde{c}_b)+\partial_{\tilde{z}}(\tilde{q}_{b,\tilde{z}} \tilde{c}_b)\nonumber\\
+\tilde{\mu}_n\tilde{\theta}_a\tilde{\rho}_a \tilde{c}_b/(\tilde{k}_n+\tilde{c}_b))d\tilde{y}&&=0.
\end{eqnarray*}
Interchanging the integration and the differentiation operators, these equations become
\begin{eqnarray}
\partial_{\tilde{t}}\bigg(\int_{-(1-\tilde{d})}^{1-\tilde{d}} \tilde{c}_wd\tilde{y}\bigg)+2\partial_{\tilde{t}}\tilde{d} \tilde{c}_w\big|_{\tilde{y}=-(1-\tilde{d})}-\partial_{\tilde{z}}\bigg(\int_{-(1-\tilde{d})}^{1-\tilde{d})}\bigg(\frac{1}{\text{Pe}}\partial_{\tilde{z}} \tilde{c}_w-\tilde{q}_{w,\tilde{z}}\tilde{c}_w\bigg)d\tilde{y}\bigg)\label{n333}\\
-2\partial_{\tilde{z}}\tilde{d}\bigg(\frac{1}{\text{Pe}}\partial_{\tilde{z}} \tilde{c}_w-\tilde{q}_{w,\tilde{z}}\tilde{c}_w\bigg)\big|_{\tilde{y}=-(1-\tilde{d})}+2\bigg(\frac{1}{\epsilon^2\text{Pe}}\partial_{\tilde{y}} \tilde{c}_w-\tilde{q}_{w,\tilde{y}}\tilde{c}_w \bigg)|_{\tilde{y}=-(1-\tilde{d})}&&=0,\nonumber\\
2\partial_{\tilde{t}}\bigg(\int_{-1}^{-(1-\tilde{d})}\tilde{\theta}_w \tilde{c}_bd\tilde{y}\bigg)-2\partial_{\tilde{t}}\tilde{d}\tilde{\theta}_w \tilde{c}_b\big|_{\tilde{y}=-(1-\tilde{d})}\nonumber\\
-2\partial_{\tilde{z}}\bigg(\int_{-1}^{-(1-\tilde{d})}\bigg(\frac{\tilde{\theta}_w}{\text{Pe}}\partial_{\tilde{z}} \tilde{c}_b-\tilde{q}_{b,\tilde{z}}\tilde{c}_b\bigg)d\tilde{y}\bigg)+2\partial_{\tilde{z}}\tilde{d}\bigg(\frac{\tilde{\theta}_w}{\text{Pe}}\partial_{\tilde{z}} \tilde{c}_b-\tilde{q}_{b,\tilde{z}}\tilde{c}_b\bigg)\big|_{\tilde{y}=-(1-\tilde{d})}\label{n22}\\
-2\bigg(\frac{\tilde{\theta}_w}{\epsilon^2\text{Pe}}\partial_{\tilde{y}} \tilde{c}_b-\tilde{q}_{b,\tilde{y}}\tilde{c}_b \bigg)|_{\tilde{y}=-(1-\tilde{d})}+2\tilde{\mu}_n\tilde{\rho}_a\int_{-1}^{-(1-\tilde{d})}\tilde{\theta}_a \frac{\tilde{c}_b}{\tilde{k}_n+\tilde{c}_b}d\tilde{y}&&=0.\nonumber
\end{eqnarray}
Next, the lower order terms in the equations for the conservation of nutrients (\ref{S11}-\ref{S12}) are 
\[\partial_{\tilde{y}}^2\tilde{c}_{w,0}=0,\hspace{1cm}\partial_{\tilde{y}}^2\tilde{c}_{b,0}=0.\]
The interface coupling condition (\ref{S13}) becomes $\tilde{\theta}_w\partial_{\tilde{y}}\tilde{c}_{b,0}=\partial_{\tilde{y}}\tilde{c}_{w,0}$ and (\ref{S14}) becomes $\tilde{\theta}_w\tilde{c}_{b,0}=\tilde{c}_{w,0}$, while the boundary condition on the wall (\ref{S14s}) becomes $\partial_{\tilde{y}}\tilde{c}_{b,0}=0$. The symmetry in $\tilde{y}$ implies that both nutrient concentrations do not depend on $\tilde{y}$, resulting in $\tilde{c}_{w,0}(\tilde{z},\tilde{t})=\tilde{\theta}_w\tilde{c}_{b,0}(\tilde{z},\tilde{t})=\tilde{c}_0(\tilde{z},\tilde{t})$. Using the aforementioned results, both equations (\ref{n333}) and (\ref{n22}) can be written as
\begin{eqnarray*}
\partial_{\tilde{t}}(\tilde{c}_0(2-2\tilde{d}_0))+2\partial_{\tilde{t}}\tilde{d}_0 \tilde{c}_0\big|_{\tilde{y}=-(1-\tilde{d}_0)}-\frac{2-2\tilde{d}_0}{\text{Pe}}\partial_{\tilde{z}}^2 \tilde{c}_0+\tilde{c}_0\int_{-(1-\tilde{d}_0)}^{1-\tilde{d}_0}\tilde{q}_{w,\tilde{z},0}d\tilde{y}\nonumber\\
-2\partial_{\tilde{z}}\tilde{d}_0\bigg(\frac{1}{\text{Pe}}\partial_{\tilde{z}} \tilde{c}_0-\tilde{q}_{w,\tilde{z},0}\tilde{c}_0\bigg)\big|_{\tilde{y}=-(1-\tilde{d}_0)}+2\bigg(\frac{1}{\epsilon^2\text{Pe}}\partial_{\tilde{y}} \tilde{c}_0-\tilde{q}_{w,\tilde{y},0}\tilde{c}_0 \bigg)|_{\tilde{y}=-(1-\tilde{d}_0)}&&=0,\\
\partial_{\tilde{t}}(2\tilde{d}_0\tilde{c}_0)-2\partial_{\tilde{t}}\tilde{d}_0 \tilde{c}_0\big|_{\tilde{y}=-(1-\tilde{d}_0)}-2\tilde{d}_0\frac{1}{\text{Pe}}\partial^2_z \tilde{c}_0+\tilde{c}_0\int_{-1}^{-(1-\tilde{d}_0)}\tilde{q}_{b,\tilde{z},0}d\tilde{y}\nonumber\\
+2\partial_{\tilde{z}}\tilde{d}_0\bigg (\frac{1}{\text{Pe}}\partial_{\tilde{z}} \tilde{c}_0-\tilde{q}_{b,\tilde{z},0}\tilde{c}_0\bigg)\big|_{\tilde{y}=-(1-\tilde{d}_0)}-2\bigg(\frac{1}{\epsilon^2\text{Pe}}\partial_{\tilde{y}} \tilde{c}_0-\tilde{q}_{b,\tilde{y},0}\tilde{c}_0\bigg)|_{\tilde{y}=-(1-\tilde{d}_0)}\nonumber\\
+2\tilde{d}_0\tilde{\mu}_n\tilde{\rho}_a\tilde{\theta}_{a,0}\frac{\tilde{c}_0}{\tilde{k}_n+\tilde{c}_0}&&=0.
\end{eqnarray*}
Then, adding both equations and using the interface condition (\ref{S13}), we finally obtain
\[\partial_{\tilde{t}}\tilde{c}_0+\partial_{\tilde{z}}(\tilde{c}_0\langle\tilde{q}\rangle-\partial_{\tilde{z}} \tilde{c}_0/\text{Pe} )=-\tilde{d}_0\tilde{\mu}_n\tilde{\theta}_{a,0}\tilde{\rho}_a\tilde{c}_0/(\tilde{k}_n+\tilde{c}_0).\]
We focus on the water-biofilm interface (\ref{S26}):
\[
\partial_{\tilde{t}} \tilde{d}=\left\{\begin{array}{ll}
f^{+}(-u_{ref}(-\tilde{u}_{\tilde{y}}+\partial_{\tilde{z}}\tilde{d} \tilde{u}_{\tilde{z}})/q_{ref}),&\qquad\tilde{d}=1,\\
-(1+(\epsilon\partial_{\tilde{z}} \tilde{d})^2)^{1/2}\epsilon \tilde{k}_{str}\tilde{\mu} \tilde{S}-u_{ref}(-\tilde{u}_{\tilde{y}}+\partial_{\tilde{z}}\tilde{d} \tilde{u}_{\tilde{z}})/q_{ref},&\qquad 0<\tilde{d}<1,\\
0,&\qquad\tilde{d}=0.\end{array}\right.
\]
Using the set-valued Heaviside graphs (\ref{hev}), we can write the previous equations as
\begin{eqnarray}\label{rober}
\partial_{\tilde{t}} \tilde{d}\in && H_0(\tilde{d})H_{1}(\tilde{d})) (-(1+(\epsilon\partial_{\tilde{z}} \tilde{d})^2)^{1/2}\epsilon \tilde{k}_{str}\tilde{\mu} ||(\mathbb{I}-\pmb{\tilde{\nu}}\pmb{\tilde{\nu}}^T)(\mathbb{\tilde{M}}+\mathbb{\tilde{M}}^T)\pmb{\tilde{\nu}}||\\
&&-u_{ref}(-\tilde{u}_{\tilde{y}}+\partial_{\tilde{z}}\tilde{d} \tilde{u}_{\tilde{z}})q_{ref})+(1-H_{1}(\tilde{d}))f^{+}(-u_{ref}(-\tilde{u}_{\tilde{y}}+\partial_{\tilde{z}}\tilde{d} \tilde{u}_{\tilde{z}})/q_{ref}).\nonumber
\end{eqnarray}
Using the regularized Heaviside functions (\ref{hevr}), we can write (\ref{rober}) as
\begin{eqnarray*}
\partial_{\tilde{t}} \tilde{d}= && H_{\delta,0}(\tilde{d})H_{\delta,1}(\tilde{d})) (-(1+(\epsilon\partial_{\tilde{z}} \tilde{d})^2)^{1/2}\epsilon \tilde{k}_{str}\tilde{\mu} ||(\mathbb{I}-\pmb{\tilde{\nu}}\pmb{\tilde{\nu}}^T)(\mathbb{\tilde{M}}+\mathbb{\tilde{M}}^T)\pmb{\tilde{\nu}}||\\
&&-u_{ref}(-\tilde{u}_{\tilde{y}}+\partial_{\tilde{z}}\tilde{d} \tilde{u}_{\tilde{z}})/q_{ref} )+(1-H_{\delta,1}(\tilde{d}))f^{+}(-u_{ref}(-\tilde{u}_{\tilde{y}}+\partial_{\tilde{z}}\tilde{d} \tilde{u}_{\tilde{z}})/q_{ref}).
\end{eqnarray*}
Using (\ref{S27}-\ref{S28}, \ref{chacab}a, \ref{velpot}a, \ref{sobres}), for the lower-order terms in $\epsilon$ we have
\begin{eqnarray*}
\partial_{\tilde{t}} \tilde{d}= H_{\delta,0}(\tilde{d})H_{\delta,1}(\tilde{d}))(-\tilde{k}_{str}(1-\tilde{d}_0)|\partial_{\tilde{z}}\tilde{p}_0|+\tilde{d}_0\tilde{\Sigma}_0)+(1-H_{\delta,1}(\tilde{d}))f^{+}(-\tilde{\Sigma}_0).
\end{eqnarray*}
Letting $\delta$ go to zero in order to return to the nonregularized formulation, we get 
\[\partial_{\tilde{t}} \tilde{d}_0=
\left\{\begin{array}{ll}
f^{-}(\tilde{\Sigma}_0) &\qquad\tilde{d}_0=1,\\
-\tilde{k}_{str}(1-\tilde{d}_0)|\partial_{\tilde{z}}{\tilde{p}_0}|+\tilde{d}_0\tilde{\Sigma}_0, &\qquad 0<\tilde{d}_0<1,\\
0, &\qquad d=0.\\
\end{array}\right.\]

\bibliographystyle{numeric}

\end{document}